\documentstyle[psfig]{texas}

\def\hmpc{{\rm \; h^{-1}\,Mpc}}
\def\kmpc{\rm \,h\,Mpc^{-1}}

\def\x{{$\xi(r)$\ }}

\def\xir{$\xi(r)$\ }
\def\xis{$\xi(s)$\ }
\def\xip{$\xi(r_p,\pi)$\ }
\def\xip{$\xi(r_p,\pi)$\ }
\def\xicc{$\xi_{cc}$\ }
\def\wp{$w_p(r_p)$\ }
\def\pk{{$P(k)$}} 

\def\kms{\,{\rm km\,s^{-1}}}

\def\kvec{{\bf k}}
\def\ss{{\bf s}\ }

\def\n_med{{\left<n\right>}}

\def\begc{\begin{center} }
\def\endc{\end{center} } 
\def\begf{\begin{figure} }
\def\endf{\end{figure} }

\def\j3{{J_3}}

\begin{document}

\title{LARGE-SCALE STRUCTURE AT THE TURN OF THE
MILLENNIUM\footnote{Invited review delivered at the 19$^{th}$ Texas
Symposium on Relativistic Astrophysics, Paris, December 1998.}
}
\author{Luigi Guzzo}
\address{Osservatorio Astronomico di Brera\\
Via Bianchi 46, I-23807, Merate, Italy\\
{\rm Email: guzzo@merate.mi.astro.it}}

\begin{abstract}
I review the current status of studies of the large--scale structure
of the Universe using redshift surveys of galaxies and clusters of
galaxies.  
I first summarise the advances we have made in our knowledge of the {\it
cosmography} of the $z<0.2$ Universe during the last 25 years,
as well as the status of the major surveys in progress.  The question
of how the {\it a priori} 
selection of some classes of objects {\it biases} the mapping of the
underlying mass density field is discussed in some detail.
I then emphasise the advantages of using
clusters of galaxies selected in the X-ray band as tracers of
large--scale structure, summarising the most recent results of the 
REFLEX survey, which is under completion.  The strong potential 
of using X-ray clusters to study the evolution of structure to large
redshifts is underlined.   
I then summarise some of the most recent statistical results on the
clustering of galaxies and clusters, using the two--point
correlation function \xis and the power spectrum \pk.  In particular,
I concentrate on the increased information available on the detailed
shape of these functions on large scales,  $\lambda\sim 100\hmpc$.  I
argue that significant evidence is accumulating from different
observations that the power spectrum has a well--defined and possibly
narrow peak around
$k\sim 0.05\kmpc$.  In the near future, measures of \pk\ from the full
REFLEX survey, from the 2dF survey, and in particular from the SDSS 
large--volume subsamples will be crucial checks for these indications.
I conclude with a glimpse into the future of large--scale structure
surveys at high redshifts, describing the features of the VIRMOS deep
survey, which will soon start collecting redshifts with the ESO VLT
for $\sim$150,000 galaxies at a typical depth of $z=1$.

\end{abstract}

\section{Introduction}

Our view of the large--scale distribution of luminous objects
in the Universe has changed dramatically during the last 25 years:
from the simple pre--1975 picture of a distribution of ``field'' and 
``cluster'' galaxies, to the discovery of the first single 
superstructures and voids, to the most recent results showing an 
almost regular web--like network of interconnected clusters,  
filaments and walls, separating huge nearly--empty volumes.  
The increased efficiency of redshift surveys, made possible by the
development of fast spectrographs and -- especially in the last
decade -- by an enormous increase in their multiplexing gain (i.e.\
the ability to collect spectra of several galaxies at once),
has allowed us not only to do {\it cartography} of the nearby
Universe, but also to statistically characterise some of its properties.
At the same time, parallel advances in the theoretical modeling of the
development 
of structure, with large high--resolution gravitational simulations
coupled to a deeper -- yet limited -- understanding of how 
to form galaxies within dark--matter halos, have provided a more
realistic connection of the models to the observable quantities.   
Despite the large uncertainties that still exist, this has transformed 
the study of cosmology and large--scale structure into a truly quantitative 
science, where theory and observations can progress side by side.

I have been asked by the organizers of the 19$^{th}$ Texas Symposium
to review this progress, and this paper is the result of this
effort.  It is clearly impossible, and actually beyond the scope of 
a review this size, to provide a thorough and complete 
summary of all the work done in the field of large--scale structure
during this intense period of growth.  There are a number
of excellent reviews that appeared in the literature in recent years,
from which the interested reader can build his/her own personal
and more comprehensive view of the historical development of this
relatively young branch of cosmology.  If I were to suggest a 
pedagogical tour on this subject, I would personally start with 
Rood \cite{rood}, who provides an enthusiastic 
first-hand description of the early pioneering years.  Within
this paper, the reader can find all the relevant references
to almost anything done before its publication.  I would then
continue with Geller \& Huchra \cite{GH89}, and Giovanelli \& Haynes 
\cite{ricmartha91}, who give a summary of
the important work done during the eighties, which saw the completion
of the CfA1 survey by Davis and collaborators, its extension (CfA2),
as well as the Perseus-Pisces \cite{ricmartha91} survey.   
More recently, Strauss \& Willick \cite{SW94} provide a
tutorial about the study of both the distribution and motions of
galaxies, while in Borgani \cite{stef95}  
one can find a thorough introduction to several statistics applied to 
the distribution of galaxies.   Further updates, including
early descriptions of projects which are also discussed here in a
more advanced stage of development, are given by Guzzo \cite{G96},  
Strauss \cite{Strauss97}, and more recently by Chincarini \& Guzzo
\cite{chin-guz} and da Costa \cite{DaCosta99}.

Here, I will try and elaborate on a few selected highlights, in the
attempt of clarifying -- or at least giving a hint of -- what we
know and what we do not seem to understand yet concerning
the properties and the origin of large--scale structure.  Emphasis
will be on the ideas, and I hope my colleagues will
forgive me if the discussion is not always as rigorous as it might
formally be.  Even if this cannot be a comprehensive review paper,
I have done my best to be as complete as possible in terms of at 
least mentioning the most relevant work done or in progress, with 
a proper link to a corresponding paper (or web page).  Conversely, I
have also tried, whenever possible, to recall the basic 
concepts required to make the discussion as self--contained as
possible.  Obviously,  the topic selection reflects my personal taste,
interests, and ignorance, so I apologise in advance to all 
colleagues whose work I might have overlooked\footnote{Although the
core of this review reflects the talk given at the 19$^{th}$ Texas
Symposium in December 1998, I found it appropriate to include also a few
references to works that appeared till March 1999.}.  

Unless differently specified, throughout the paper the Hubble constant 
will be parameterised as H$_o=100/h\, \kms$ Mpc$^{-1}$, and a model with 
q$_o=0.5$ and $\Lambda=0$ will be adopted.

\section{The Large--Scale Galaxy Distribution within $z\le 0.2$}

How well do we know the global picture, the
``geography'' of large-scale structure?   At the very basic level, any theory
of large-scale structure and galaxy formation must be able to reproduce the
visual appearance of the large-scale galaxy distribution.  For this reason
the first result of a redshift survey is essentially a map of the galaxy
distribution in space.  More than for the sometimes complicated statistical 
analyses that can be computed from them, these maps are what 
redshift surveys become usually famous for (many of us certainly remember
the Coma region ``homunculus" of the first CfA2 slice \cite{delapp86}).
For this reason and possibly because this responds to the need of 
human beings to ``see'' where they stand within the Universe, we like so 
much to produce ``cone'' or ``wedge''  diagrams:  something
that was just a picture over the sky vault, for the first time is seen in
its almost real spatial distribution thanks to the newly measured distances.
Such a feeling has probably something in common with that of ancient 
explorers, when they discovered a new piece of territory not previously
marked on the maps.

\subsection{From 2D Photometric Galaxy Catalogues to Redshift Surveys}

However, there is at least one major difference to this romantic view.
In fact, when we start a 
redshift survey, we inevitably have to know in advance that a galaxy is 
there, with its right ascension, declination and apparent magnitude or 
flux in some band of the electromagnetic spectrum.   A necessary 
prerequisite for any redshift survey is, in other words, the availability
of a {\it photometric catalogue}, from which an apparent--magnitude
(or flux) limited sample can be extracted.   This introduces an unavoidable
prejudicial selection effect that determines {\it how} our survey 
will trace the large--scale distribution of matter.  In this paper, 
we shall come 
back several times to this concept, which is becoming more and more
important now that large redshift surveys are starting to explore
depths to which the issues of galaxy evolution can no longer be
reasonably ignored.

\begin{figure}
\centering
\psfig{file=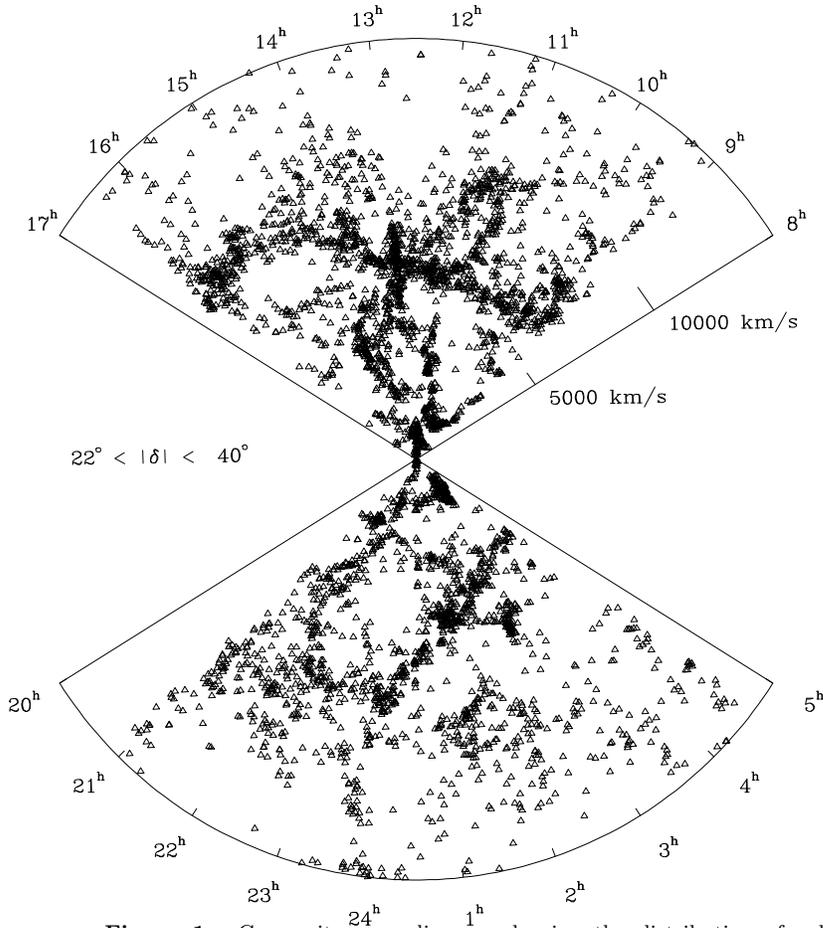,width=13cm}
\caption{Composite cone diagram showing the distribution of galaxies
with $m_B< 15.5$ in a $18^\circ$ degree thick slice from the combined CfA2 
and SSRS2 redshift surveys (from \cite{DaCosta99}).  The structure spanning the
whole aperture of the Northern cone is the famous {\it Great Wall}.}
\label{cfa-ssrs}
\end{figure}

The CfA1 \cite{Davis80}, CfA2 \cite{GH89} and
Perseus--Pisces \cite{ricmartha91} redshift surveys,
for example, were all possible in their era because Fritz Zwicky 
and collaborators had previously
constructed a catalogue of galaxy positions and magnitudes
down to a photographic magnitude $m_{ph}\simeq 15.7$ over the whole
Northern hemisphere \cite{Zwicky}.   When, for example, 
the Southern Sky Redshift Survey (SSRS, to $m_B\simeq 14.5$ and SSRS2, 
to $m_B\simeq 15.5$, see \cite{DaCosta99}) --
which aimed at being the southern equivalent of the
CfA survey -- was started, a major difficulty involved the
construction of a ``quasi-Zwicky'' magnitude--limited sample
homogeneous to the northern one from the catalogues available in the
South (essentially the ESO--Uppsala diameter--limited 
catalogue, \cite{Lauberts}) \footnote{In a similar fashion, the first
attempt to extend the CfA2 survey to fainter magnitudes ($r=16.1$),
with the 1--degree ``Century Survey''\cite{century}, required the scanning and
calibration of red POSS--E plates to build the parent photometric
catalogue.}.   The matching of the two surveys 
(CfA2 and SSRS2), produced a 
sample of more than 15,000 galaxies with measured redshift which still is the
most representative description of the details of the large--scale 
distribution of bright galaxies within the local
$r \sim 100\hmpc$ volume.  This can be appreciated from the cone
diagram of Figure~\ref{cfa-ssrs}, where a $18^\circ$ slice 
through these combined data is reproduced from \cite{DaCosta99}.  

The cosmological importance of large, homogeneously--selected photometric
catalogues of galaxies was certainly first appreciated (or at least first
translated into a concrete effort), by the British astronomical
community\footnote{In fact, at the same time considerable effort was
spent in a similar direction by the M\"unster group in
Germany. This was limited to a smaller area, and used objective--prism
plates to measure very low--resolution redshifts for nearly a million galaxies
\cite{MRSP}.}.  
At the end of the eighties two different groups in the UK 
started independently to scan, analyse and
calibrate through dedicated CCD photometry the IIIa-J plates 
of the UK--Schmidt survey.  The two projects used
the Automated Plate Measuring (APM) Machine in Cambridge and the COSMOS
machine in Edinburgh, respectively, and their final products are represented 
by the well--known APM galaxy catalogue \cite{APM90} and
Edinburgh--Durham Southern Galaxy Catalogue (EDSGC), \cite{EDSGC}.
The realization of a similar catalogue in the Northern 
hemisphere has not been possible till the present, due to the lower depth of the
Palomar survey plates, but is now becoming a reality with the completion
and analysis of the deeper POSSII survey \cite{Djorgowski98}.

The APM catalogue covers 1.3 sr in the Southern hemisphere (185 Schmidt
plates), while the EDSGC is limited to the 60 Schmidt plates around the South
Galactic Pole.   Both catalogues
are complete to $b_J\simeq 20.5$, i.e. about 5 magnitudes deeper than the
Zwicky catalogue (whose photometric band is not too dissimilar to
the $b_J$), and still represent excellent and not yet
exploited databases for fairly deep, wide--area redshift surveys.

Not only are 2D galaxy catalogues fully important as source lists 
for redshift surveys, but they also bear significant statistical
information on large--scale structure studies {\it per se}.
One important example of the results obtained directly from the two
digitised catalogues discussed above is the angular correlation
function $w(\theta)$ \cite{APM90,CollinsWtheta}.  These measures were
largely responsible for killing the once fashionable standard CDM
model (where by ``standard' one meant $\Omega=1$, $H_o=50$ km s$^{-1}$
Mpc$^{-1}$, and a bias parameter $b=2.5$ \cite{SCDM} -- see
\S~\ref{sec:bias} for definitions --).    
First redshift surveys based on the two catalogues started relatively early,
in particular {\it sparsely sampled} surveys, which represented a compromise
between the wish to exploit the large areas available and the amount of 
telescope time needed to cover them spectroscopically.   
This is the case of the Stromlo--APM survey 
and the Durham-UKST survey, from which scientific
results have been produced until very recently.  

The Stromlo--APM survey includes 1797 redshifts, for galaxies selected
-- one out of every twenty -- over the whole APM catalogue down to
$b_j=17.15$, observed using traditional  
single--slit spectroscopy (see e.g. \cite{Stromlo-APM}).
The Durham--UKST survey instead, measured redshifts for $\sim 2500$ galaxies,
selecting one in three objects with magnitude brighter than $b_j\simeq
17$ from the smaller area of the EDSGC (see e.g. \cite{Ratcliffe et
al. 1996}).  This survey was constructed using FLAIR at the UK Schmidt
Telescope, which is a fibre optic system capable of collecting the
light of $\sim 50$ galaxies over the area of a Schmidt plate and
bringing it into the slit of a conventional spectrograph standing on
the dome floor \cite{flair}.

\subsection{The Era of Multi--Object Spectroscopy}

Indeed, one very specific aspect that made these 2D catalogues particularly 
attractive for redshift surveys, was the parallel development of
fibre--optic spectrographs, capable of observing  several tens of
galaxy spectra over typical areas of half a degree in diameter.
The first successful examples of these instruments were 
mounted at the Cassegrain focus of 4~m class telescopes:  following 
the pioneering work of the eighties (see \cite{Hill} for a review),
relevant examples at the time were FOCAP and then AUTOFIB at the
Anglo--Australian Telescope, and OPTOPUS at the ESO 3.6~m telescope
\cite{Avila89}.    The density of
fibres on the sky provided by these instruments corresponded to around 
100-200 deg$^{-2}$.  These figures are matched by the average number
counts of galaxies on the sky for blue limiting magnitudes $\sim
19-19.5$, i.e. well within the limits of the EDSGC and APM catalogues.

\begin{figure}
\centering
\psfig{file=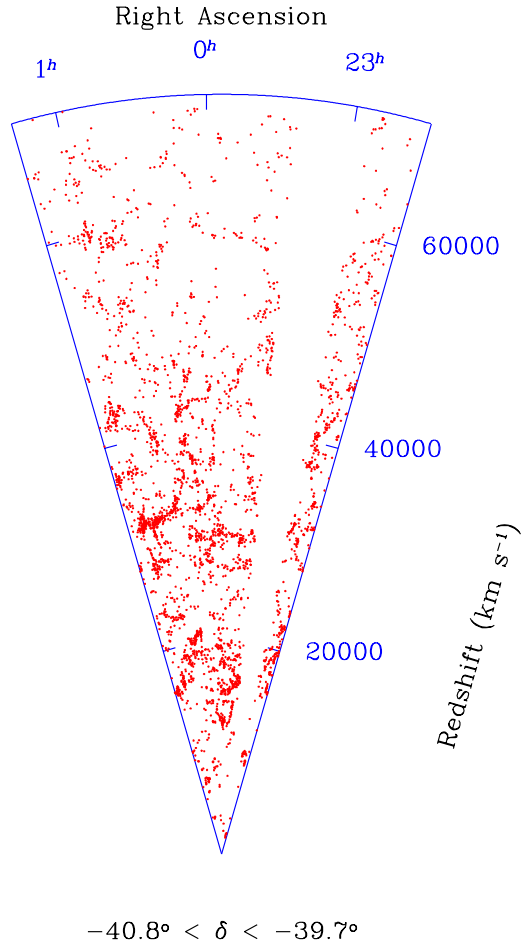,width=13cm}
\caption{The distribution of galaxies within the 
1-degree slice of the ESP survey (the gap around $23^h$ corresponds to
a region that was not observed).}  
\label{esp-cone}
\end{figure}
This reasoning was at the origin of the ESO Slice Project (ESP)
redshift survey, which used the ESO Optopus fibre coupler to observe
all objects in the EDSGC down to $ b_J= 19.4$ within a thin strip of
$\sim 1^\circ \times 26^\circ$, constructing an 85\% redshift--complete
sample of 3348 galaxies.  The step forward in depth was gigantic, with
respect to the Zwicky--based surveys, nearly 4 magnitudes, which
lead to an effective survey depth\footnote{The survey depth is
defined as the maximum distance to which a galaxy of magnitude M$^*$, i.e.
the characteristic magnitude in the Schechter form \cite{PS76} for the 
luminosity function, is detected within the survey.}
corresponding to a luminosity distance of $\sim 600\hmpc$.
A pictorial view of the large-scale distribution of
galaxies in the ESP is shown in the cone diagram of
Figure~\ref{esp-cone}. The survey and the main results obtained from
its analyses have been largely discussed in a series of papers
\cite{Vett97,Zucca97,Vett98,Cappi98,Scaramella98,Ramella98,Guzzo99}.
I shall summarise the main results obtained on 
galaxy clustering in the ESP survey in \S\ref{sec:stat}.

Also based on intensive use of a multi--object fibre spectrograph, is
the largest redshift survey completed to date, i.e. the Las Campanas
Redshift Survey (LCRS \cite{LCRS96}), which represents the best
existing compromise between depth and angular aperture.  This redshift
catalogue is slightly less deep than the ESP ($r<17.7$, roughly
corresponding to $b_J\sim 18.8$ for a typical mean galaxy colour), 
but covers six slices of $\sim 80^\circ \times 1.5^\circ $, for a
total of $\sim 26,000$ redshifts.   Unlike most of the aforementioned
surveys, the photometric parent sample for the LCRS was constructed
anew from a specifically performed CCD drift--scan survey in the $r$
band.  
\begin{figure}
\centering
\psfig{file=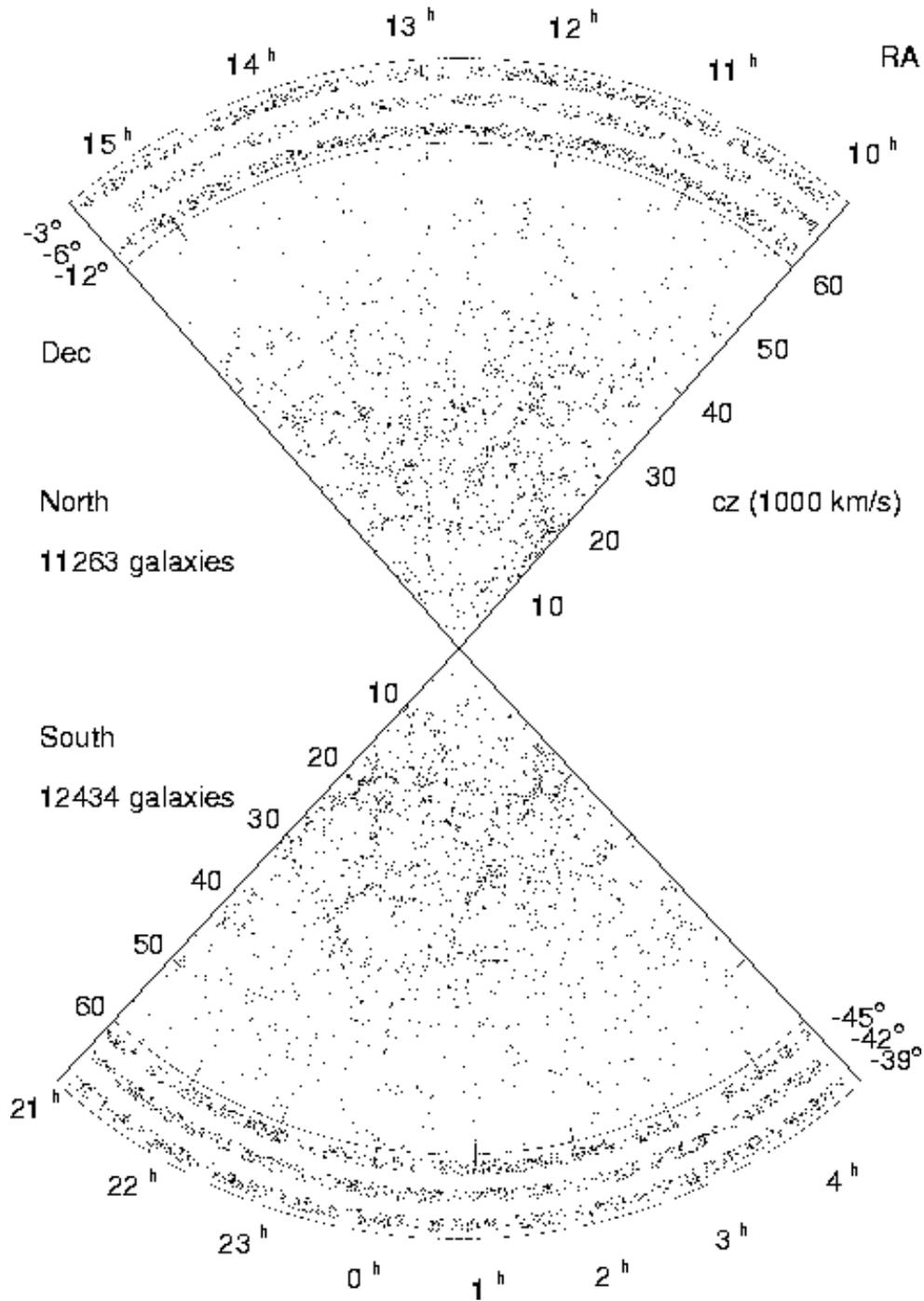,width=13cm}
\caption{The famous double--cone diagram from the Las Campanas Redshift
Survey.  Both cones  show galaxies projected from three adjacent RA slices of
$1.5^\circ$ each \cite{LCRS96}.}
\label{lcrs_cones}
\end{figure}
One peculiarity of the LCRS is that the selection of the target
galaxies was subject not only to a cut in apparent magnitude, but
also to a selection in surface brightness within the aperture
of the used fibres.   It is now clear (Dalcanton, private comm.),
that this selection favours bulge--dominated galaxies, preferentially
excluding the irregular galaxies that represent the main population 
at faint luminosities.  This has a clear effect on the galaxy
luminosity function as measured from the LCRS, showing up in
an abnormally flat faint end \cite{LCRS_LF}, (see \cite{Zucca97} for
comparison to other surveys).    
On the other hand, the global clustering properties do not seem
to be significantly affected by this selection, as judged from the comparison
of its two--point correlation function with those of the ESP and
Stromlo--APM surveys (\cite{Guzzo99}, see also Figure~\ref{xi-surveys} 
hereafter).   The double cone diagram showing the large--scale
distribution of galaxies in the six LCRS slices is reproduced in
Figure~\ref{lcrs_cones}.  

\begin{figure}
\centering
\psfig{file=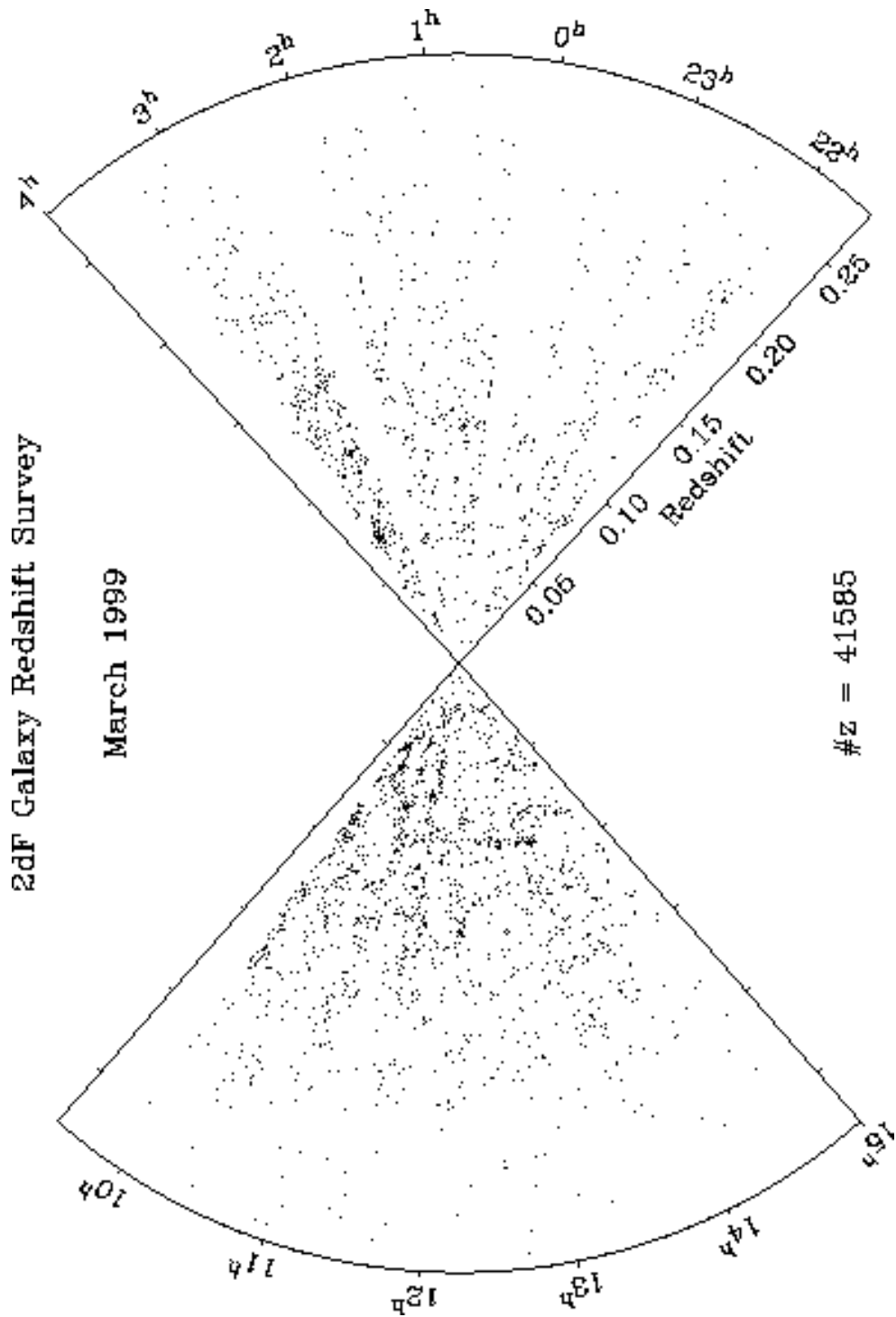,width=13cm}
\caption{The distribution of the 41585 galaxies observed so far (March
1999)in the 2dF survey: like light beams through a smoky room, the 2dF
observations are steadily unveiling the details of large--scale
structure, at a typical rate of $\sim 2000$ redshifts per observing
night. (Courtesy of the 2dF team).
}
\label{2dF-cone}
\end{figure}
The advantages of multi--object fibre spectroscopy have been 
pushed to the extreme with the construction of the 2--degree--Field (2dF)
spectrograph for the prime focus of the Anglo--Australian Telescope
\cite{2dF}. This instrument is able to accommodate 400 automatically
positioned fibres over a 2--degree--diameter field.  This implies
a density of fibres on the sky of $\sim 130$ deg$^{-2}$, and an optimal
match to the galaxy counts for a magnitude similar to that of
the ESP survey, $b_J\simeq 19.5$.   The striking difference is that
with such an area yield, a number of redshifts as in the ESP survey
(although not distributed over a strip) can be collected in $\sim 10$
exposures, i.e. slightly more than one night of telescope time with typical 1
hour exposures!  It was rather natural, therefore, that a large
redshift survey based on the 2dF spectrograph were proposed by a
UK-Australian team.  This survey is now known as the 2dF galaxy redshift
survey, and is based on the APM catalogue, giving us one
further example of how importance of such large
photometric catalogues.  Its goal is to measure redshifts for more
than 250,000 galaxies with $b_J\le 19.5$. About 4/5 of these lay 
within two large areas, $75^\circ \times 12.5^\circ$ and $65^\circ
\times 7.5^\circ$ within the South and North Galactic Caps
respectively, which are being fully covered with a honeycomb of 2dF
fields.  Another 40000 redshifts will be measured within 100
fields  randomly distributed over the APM area, with the goal of
maximising the signal in the power spectrum estimate on very large
scales \cite{Kaiser86}.   In addition, a faint redshift survey of   
10,000 galaxies brighter than $R=21$ will be performed over selected
fields within the two main strips.  The survey is steadily collecting 
redshifts, and first results on the luminosity function for different
morphological types (defined through their spectral properties), have
been recently presented \cite{2dF_first}.  From the cone diagram of
Figure~\ref{2dF-cone}, we can have a visual impression of the status of the 
survey as of March 1999, with a total of 41585 redshifts measured,
(note that the thickness of the slice is not uniform over the RA
range).  With the eye trained by the 
ESP and LCRS diagrams, one can easily see the structures taking shape
across the survey beams.  More details can be found in
\cite{Colless_2dF}, and \cite{Maddox_2dF}.  See also the 2dF web page at
{\tt http://msowww.anu.edu.au/$\sim$colless/2dF/}, where the diagram
of Figure~\ref{2dF-cone} is continuously updated.

The most ambitious and comprehensive galaxy survey project currently
in progress is without any doubt 
the Sloan Digital Sky Survey (SDSS).  This massive effort is carried on
by an American consortium with the participation of Japan.  Aim of the
project is first of all to observe photometrically the whole northern
galactic cap, $30^\circ$ away from the 
galactic plane ($\sim 10^4$ deg$^2$) in five bands,  at limiting magnitudes, 
respectively of $u^\prime=22.3$, $g^\prime=23.3$, $r^\prime=23.1$,
$i^\prime=22.3$ and $z^\prime=20.8$.  The expectations are to detect $\sim
5\times10^7$ galaxies and $\sim 10^8$ star--like sources among which a
subset of 
$10^6$ AGN candidates can be selected by colour techniques.   This has
already led, based on the first few hundred deg$^2$ covered since first
light (May 1998), to the discovery of several high--redshift ($z>4$)
quasars, including the highest--redshift quasar known, at
$z=5.0$ \cite{SDSS_QSOs}.   Using two fibre spectrographs carrying 320
fibres each, 
the spectroscopic part of the survey will then collect spectra for
about $10^6$ galaxies with $r^\prime < 18$ and $10^5$ AGN's with $
r^\prime < 19$.  The spectrographs are still being commissioned at the
telescope at the time of writing this paper (Spring 1999).

\begin{figure}
\centering
\psfig{file=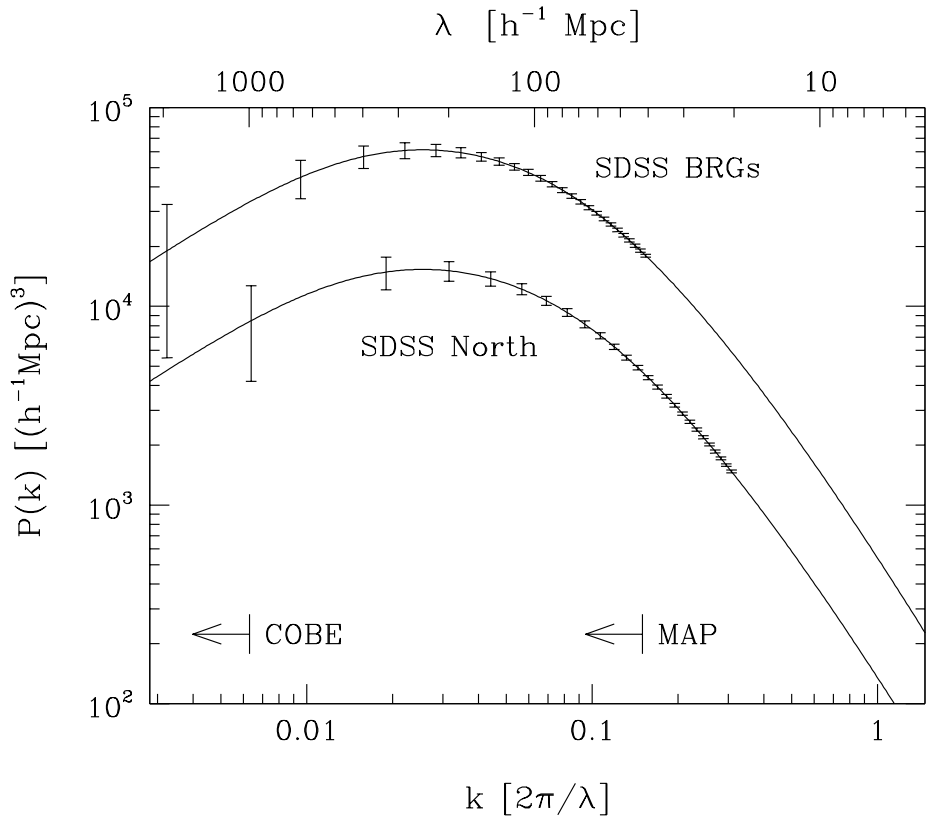,width=13cm}
\caption{Expected $1\sigma$ accuracy in the galaxy power spectrum measured from
the SDSS data.  The solid lines give the (input) power spectrum.  The bottom
curve is for the 900,000 galaxies expected in the main survey volume to
$r^\prime<18$, while the top curve is for the deeper subset of 100,000 red
luminous galaxies with $r^\prime<19.5$ (Loveday 1998). 
}
\label{sdss-pk}
\end{figure}

The capability to isolate photometrically sub--classes of objects
through their colours will be exploited also by selecting a sample of
about $10^5$ ``red'' luminous galaxies with 
$r^\prime<19.5$.  These will be observed spectroscopically providing a
nearly volume-limited sample of early--type galaxies with a 
median redshift $z\simeq 0.5$, that will be extremely valuable to study the
evolution of clustering.  In Figure~\ref{sdss-pk}, using a numerical
simulation, the expected power spectrum  
of the whole SDSS spectroscopic galaxy survey (bottom line)
is compared to that corresponding to the subsample of red
luminous galaxies \cite{Loveday98}.  Note the gain in the
clustering signal around the turnover in the power spectrum\footnote{We
shall define and discuss in more detail the power spectrum of
clustering in \S\ref{sec:pk}.}.

It is clear from this short description how the goals of the SDSS are well
beyond the pure realisation of a redshift survey for just studying large--scale
structure.   Indeed, in addition to  the immediate scientific results that will
be produced by the survey team, it will build an unparalleled photometric and
spectroscopic data base whose impact on the scientific community will
last for several years.   Further details can be found in
\cite{Margon98}, with a more recent update  as of fall 1998, in
\cite{Loveday98}.  See also the SDSS web site at {\tt
http://www.sdss.org/}, where the latest news on the ongoing survey can
be found. 

\section{A Fair Sample of the Universe?}
\label{sec:variance}

A general impression one could draw from the cone diagrams of the ESP
and LCRS surveys, and which is also taking shape within the 2dF
preliminary plot, is that inhomogeneities in the  distribution of
galaxies are limited to scales of $100-200 \hmpc$, which are fairly well
covered by these modern surveys.  Using Bob Kirshner words, we seem
to be finally seeing ``the end of greatness'', that is, we are finally
sampling (at least in two dimensions) sizes which contain the
relevant clustering scales of our Universe.   This was not the case
until a few years ago, when any new survey used to discover larger
and larger structures.  A clear example of this situation is provided
by the combined CfA2--SSRS2 sample that we have shown in
Figure~\ref{cfa-ssrs}.  Here the largest superclusters have sizes
$\sim150 \hmpc$, comparable to the survey depth, and spanning its 
volume from one side to the other.

It was because of evidences like these, together with the power--law
behaviour shown by galaxy clustering (see \S\ref{sec:xi}), that some
researchers suggested that the large--scale galaxy distribution of
galaxies could be described as a ``fractal dust''
(e.g. \cite{Mandelbrot}).  This fractal behaviour, if extrapolated to
indefinite  
scales, i.e. if not up--bounded by a transition
to a homogeneous distribution,  has some rather dramatic consequences
on our statistical description of large--scale structure
\cite{Pietronero87}.  No mean density can be defined, and as a
consequence the whole concept of {\it density fluctuations} (with
respect to a mean density) becomes nonsensical.  

A quite intense debate has developed in the last few years about
whether the available surveys really provide evidence for homogeneity
on the largest scales explored, with -- as it often happens -- a
tendency of the opposing views to crystallize on polarized positions.
As an attempt to clarify a bit the situation by addressing this very
basic question in a possibly objective way, in \cite{Guzzo97} I reviewed
some of the best available redshift survey data from this point of
view.  The application of some simple counting
statistics\footnote{In particular, the growth of the number of objects
$N(R)$ as a function of the distance $R$ from the observer for a
volume--limited sample extracted from a given survey.}, corroborated
also by other more detailed analyses \cite{Borgani94,Scaramella98},
seemed to show a general convergence to a homogeneous distribution, rather
than supporting a fractal behaviour to the largest explorable scales.
At the same time, however, it was clear (as previously remarked
\cite{G91}), that the clustering of galaxies on small and
intermediate scales could be described by  power--law ranges
consistent with a fractal, scale--free distribution.   Similar
conclusions have been reached more recently by Martinez \cite{Martinez99}.

\subsection{Density Fluctuations and Variances}

Once we are convinced that indeed a transition to homogeneity does
exist on some large scale around $\sim 100-200 \hmpc$ so
that we can define a mean density of the Universe in a sensible way,
we can quantify the distribution of objects and matter in terms of
{\it density fluctuations}. 
Given a mean density $n$, and a spherical volume of radius $R$, we can
measure the fluctuation $\delta N/N$ in the number of objects within
such spheres, and compute the {\it variance} $\sigma^2_{gal}(R)$ of
this quantity.
We shall then find that for typical optically--selected samples
this is a decreasing function of $R$ (galaxies are indeed clustered!),
with $\sigma^2_{gal}\simeq 1$ for $R=8 - 10 \hmpc$, depending slightly on the
mean luminosity of the objects considered.  In particular,
modern redshift surveys, with typical sizes exceeding a few $100\hmpc$,
are probing volumes of the Universe over which $\sigma^2_{gal}$
is significantly smaller than unity.

Unless mass is more clustered than light\footnote{It is more
naturally expected, as we shall discuss in \S\ref{sec:bias}, that
light is either clustered as mass or to some degree more clustered.
However, one could in principle also conceive scenarios in
which galaxy formation is suppressed by some physical mechanism in 
high--density regions, so that galaxies would look more homogeneously
distributed than the real mass density field.}, this should  be
reflected by a similar or even smaller variance $\sigma^2_\rho$ in the
mass density fluctuation field $\delta\rho/\rho$.  When $\sigma_\rho<1$, 
linear perturbation theory can be applied and comparison of models to
observations becomes easier.  In fact, if clustering is driven by
gravitational instability, the amplitude of perturbations in the linear regime 
grows with time in a way which is independent of the spatial
wavelength of the perturbation itself (see e.g. \cite{Paddy},
\cite{Col-Luc}, \cite{JAP}).  
For this reason, the evolution of any statistics describing the 
{\it distribution} of amplitudes at different wavelengths, as is
the case for the power spectrum \pk, and to some extent also 
of its Fourier transform the two--point correlation function \xir,
will be described by a simple growth in amplitude, without any change
in the shape.  

This means in practice that if we are able to measure
accurately the present shape  of $P(k)$ or \xir on scales where this
behaviour still holds, we have a direct probe of the initial
distribution of fluctuations, which can be directly compared to
the linear power spectra predicted by the different models.  This is
one of the main motivations for extending redshift surveys to
larger and larger volumes of the Universe.

\subsection{Mapping Light, Mapping Mass}
\label{sec:bias}

In the previous paragraph, we have quickly touched on the possibility
that the variances we observe in the distribution of light and mass 
are not strictly the same.   In fact, as we discussed, whatever
redshift survey we are performing, we are not mapping the
distribution of mass in the Universe, but rather the distribution
of objects that can be ``seen'' in some band of the electromagnetic
spectrum and that serve as possible {\it tracers} of matter distribution.
We have seen, for example, 
that the LCRS and ESP are selected in different photometric bands,
red and blue respectively, with a further important cut in
surface brightness for the LCRS, and how these differences affect measured
properties as the luminosity function \cite{Zucca97}.

While the LCRS and ESP galaxies do not show significant variations in
the global clustering properties (i.e., they apparently trace the
density field in similar ways), more serious differences are found
for example between optically--selected and infrared--selected
samples, as notably shown by the large surveys based on the IRAS
satellite infrared survey\footnote{For reasons of
space, I will not discuss in detail here the IRAS--based
redshift surveys, like the 1.2 Jy, QDOT and PSCz redshift surveys, a part
from mentioning some of the important results produced from them.
Details on the first two can be found in \cite{SW94}, while
for the most recent PSCz, see \cite{PSCz}.}.
A selection based on the IRAS infrared flux favours star-forming
galaxies and avoids rich clusters.  Consequently, 
the IRAS density field is smoother than that generally seen by
optically--selected galaxies, with a smaller correlation length ($\sim
3\hmpc$ vs. $5 \hmpc$).  Does this mean that IRAS galaxies trace
the mass density field, i.e., that they are {\it unbiased} objects?
The only way to answer this question is to compare the galaxy
distribution to the mass distribution derived independently from
dynamical observations of the peculiar velocity field\footnote{One
new powerful method for recovering the true mass distribution is
provided by the weak lensing distortions induced by large--scale
structures on the background galaxy images (e.g. \cite{Jain}).}
(see e.g. \cite{Dekel98}) and for the case of IRAS galaxies, the answer is
probably yes.  

\begin{figure}
\centering
\psfig{file=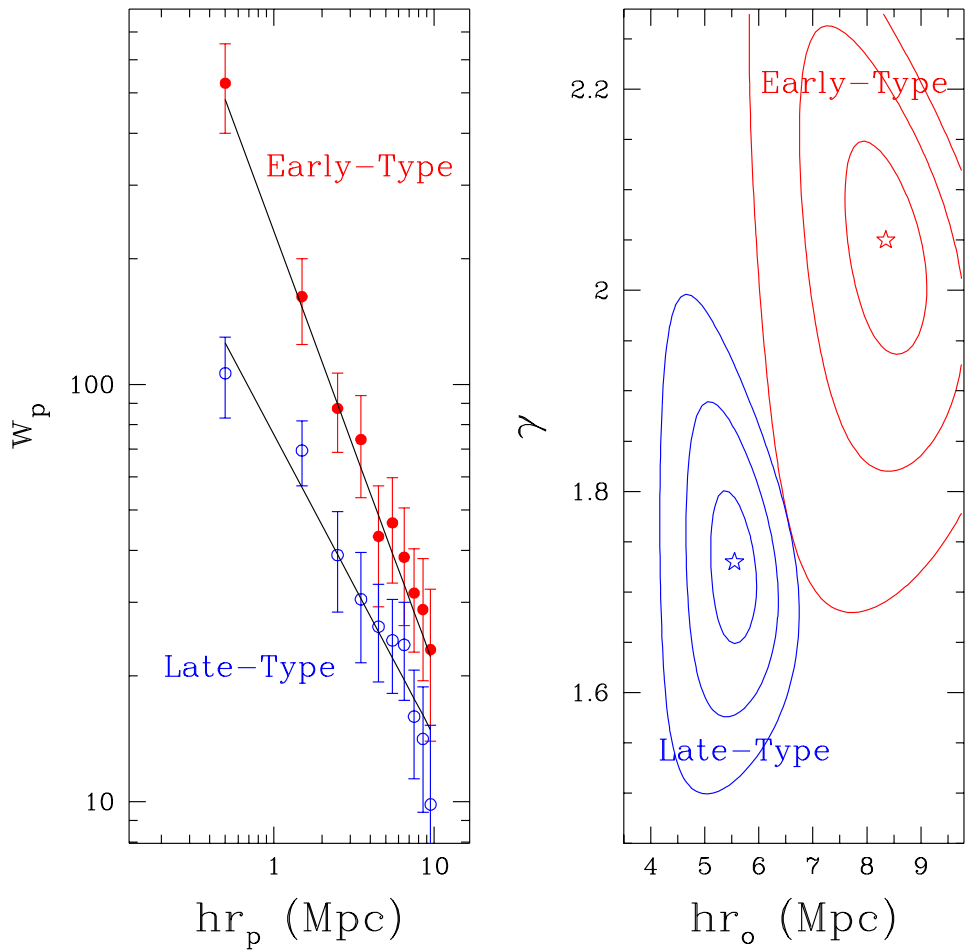,width=13cm}
\caption{Comparison of the spatial correlation functions of early-- and
late--type galaxies. The left panel shows the so--called projected correlation
function $w_p(r_p)$, i.e. a projection of galaxy correlations perpendicular to
the line of sight which has the advantage of being free from the distortions
induced by peculiar velocities (see  
e.g. \cite{G97} for details). The right panel gives the confidence
ellipses for the values of the correlation length $r_o$ and slope $\gamma$, 
when a power-law spatial correlation function $\xi(r)=(r_o/r)^{\gamma}$ is
fitted to the data.  Clearly, early--type galaxies (ellipticals and S0's) are
significantly more clustered than late--types (spirals and irregulars).  So,
which sort of galaxies (if any)  
are really tracing the large--scale mass density field?}
\label{pp_xi_types}
\end{figure}

To further clarify what we mean by {\it biased} tracers, let us have a
look at the distribution of different
morphological types.  It is well known that different galaxy types 
find themselves preferentially within different density regimes:
elliptical and S0's favour high--density regions, while spirals are
much more common in low--density environments.   This is the so--called
{\it morphology--density relation} (e.g. \cite{Dressler80}), and it inevitably
affects the way in which different types trace the underlying mass
density field.  In Figure~\ref{pp_xi_types} I have included
an estimate of the real--space two--point correlation function
\xir for early-- and late--type galaxies, 
obtained from the projected function \wp \cite{G97}.
The plot shows how the morphology--density relation translates into a stronger
clustering for elliptical galaxies.  So, if we were for some reason
able to detect only elliptical galaxies and assumed that they trace
the mass, we would at first glance conclude that matter in the
Universe is on the average more clustered than it really is.   
%

All these examples pertain to the grand challenge of understanding how
radiation emitted by cosmic objects (at any wavelength) is related to
mass, i.e. the so--called {\it bias}, and even more importantly how
this relation evolves with cosmic time.   One way to define this is to
write that
\begin{equation}
\sigma_{gal}(R) = b\, \sigma_\rho(R)\,\,\,\,\, ,
\label{sigma}
\end{equation}
where the linear bias $b$ is in general a function of scale $R$ and 
time $t$.   A proper comprehension of the behaviour of $b(r,t)$ is 
becoming crucial now that observational data on clustering at very
different cosmic epochs are being accumulated (e.g. \cite{Steidel98}).
For this reason, considerable energy has been spent in the last couple
of years in developing physically motivated bias models
(e.g. \cite{Lauro, Dekel98, Tegmark_Peebles}).  

\section{Clusters of Galaxies as Tracers of Large--Scale Structure}
With mean separations $>10\hmpc$,
clusters of galaxies are ideal objects for sampling efficiently
long--wavelength fluctuations over large volumes of the Universe.
Furthermore, fluctuations in the cluster distribution are amplified
with respect to those in galaxies, i.e. they are 
{\it biased} in much the same way as we were discussing previously
some classes of galaxies are: rich clusters form at the peaks of the
large--scale density field, and their variance is amplified by a factor
that depends on their mass, as it was first shown by Kaiser \cite{Kaiser84}.
In the next section we shall see explicitly this
effect at work in recent analyses of the clustering of clusters.

A thorough review of the use of clusters as tracers
of large--scale structure has been given recently by Postman \cite{Postman98},
with particular attention paid to optically--selected clusters.
For this reason, I will not discuss here the important issue of selecting
clusters of galaxies in the optical band, i.e. from the 2D distribution
of galaxies on the sky, but I will concentrate on results from X-ray 
selected cluster samples. Reference to optically--selected samples
will be limited to a discussion of the clustering results obtained from them.

Studies of X-ray selected clusters date back to the beginning
of X-ray astronomy \cite{Giacconi}.  However, only in recent years
statistical studies became feasible, although still limited
to the study of the temperature and luminosity functions, in
particular through the data from the Einstein Medium Sensitivity Survey \cite{Henry_Arnaud, EMSS}).   The ROSAT satellite, launched 
in 1990, not only produced serendipitous samples of distant clusters
to update these studies (see \cite{Rosati98} for a review), but
carried out the first all--sky 
survey ever with an X-ray imaging telescope.  This has  
represented a tremendous input for studies of large--scale structure
using clusters of galaxies, as I shall describe in this section.

\begin{figure}
\centering
\psfig{file=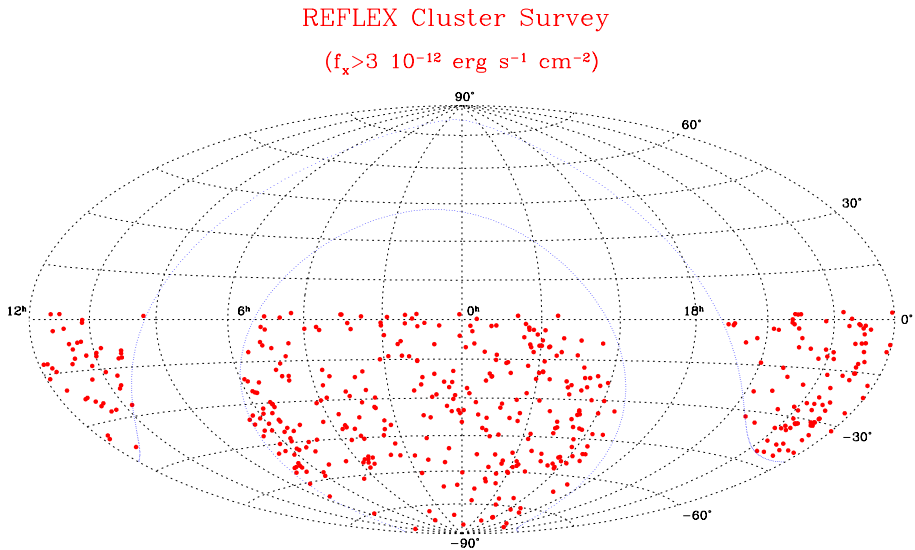,width=13cm}
\caption{The distribution on the sky of the 460 X-ray clusters in the
REFLEX survey, to $f_x> 3 \times 10^{-12}$ erg s$^{-1}$ cm$^{-2}$.}
\label{aitoff_kp}
\end{figure}
The ROSAT All-Sky Survey (RASS) was performed in the energy band
between 0.1 and 2.4 keV with the PSPC, a photon counter with a $\sim
20$ arcsec resolution on axis, degrading to nearly 2 arcmin at the
edges of the 2--degree field of view \cite{Voges_RASS}.  The median 
exposure time in the RASS was of $\sim 300$ s, which
translates in an effective detection limit for clusters of $\sim 10^{-12}$
erg s$^{-1}$ cm$^{-2}$.   Considering that in the ROSAT hard band
(0.5--2.0 keV), $L^*\simeq 1 \cdot
10^{44}$ h$^{-2}$ erg s$^{-1}$ \cite{DeGrandi_XLF}, this detection
limit translates into a typical 
depth of $d_L\simeq 900\hmpc$, i.e. $z\simeq 0.3$.  
The RASS, therefore, provides a unique opportunity to
detect clusters of galaxies within a huge volume of the ``local'' Universe.

In fact, follow--up work to construct X-ray cluster samples from the RASS
and measure their redshifts started early after completion of the
survey (see \cite{Hans_review} for a review).
A first example was a survey in the SGP area, that constructed a
sample of about 200 clusters with the main aim of measuring the
cluster--cluster 
correlation function \cite{Romer_SGP}.  Only recently, however,
the all--sky coverage of the RASS was properly exploited.  This has
been the aim of the ROSAT-ESO Flux Limited X-ray (REFLEX) cluster
survey, that uses clusters of galaxies to explore, in the Southern
hemisphere, a volume of the Universe comparable to that of the SDSS. 

\subsection{The REFLEX Survey}
\label{sec:reflex}

The REFLEX cluster survey combines the X-ray data from the RASS and optical
follow--up observations using the ESO telescopes, to construct a complete
sample of about 700 clusters with measured redshift, to a flux limit
$f_x \simeq 1.5 \times 10^{-12}$ erg s$^{-1}$ cm$^{-2}$ in the ROSAT band
(0.1--2.4 keV).
The survey covers essentially the southern celestial
hemisphere ($\delta<2.5^\circ$), at galactic latitudes 
$|b_{II}|>20^\circ$, to avoid regions of high absorption and
crowding by stars. 
\begin{figure}
\centering
\psfig{file=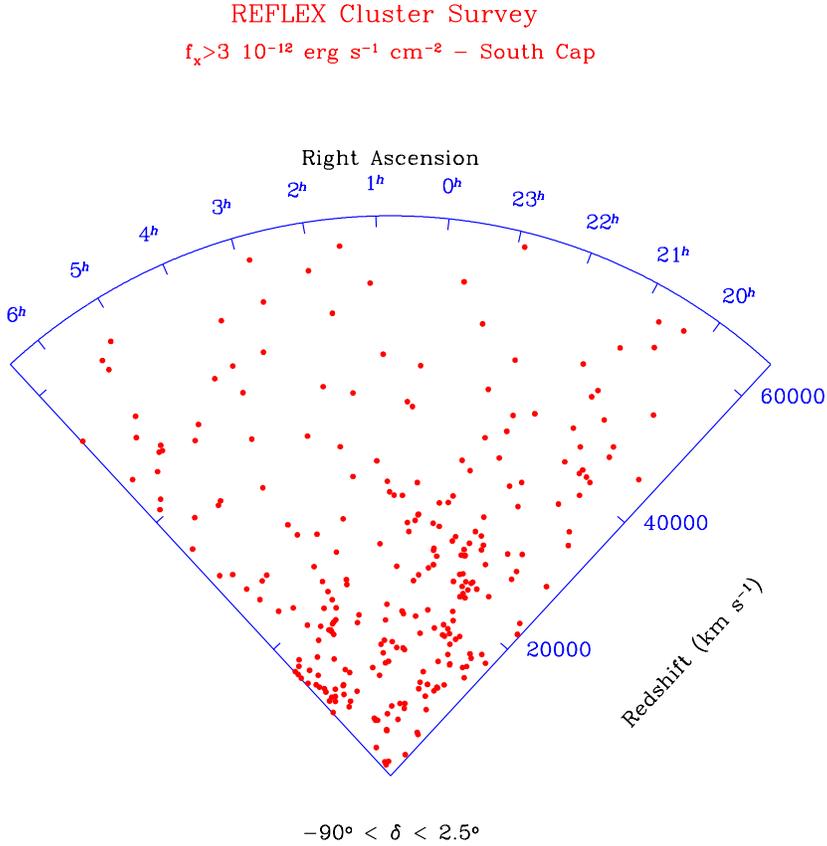,width=13cm}
\caption{The large--scale distribution of X--ray selected clusters in
the REFLEX survey.  Only the South Galactic Cap part of the survey is
shown here.
}
\label{cone_kp}
\end{figure}
During the development of the survey, this project already produced a
first bright sample of 135 clusters (the RASS1 bright sample
\cite{RASS1}), limited to the SGP area, that served as a pilot work to
fine--tune the strategy.  
At the time of writing (Spring 1999), the first REFLEX sample of nearly 460
objects with $f_x > 3 \times 
10^{-12}$ erg s$^{-1}$ cm$^{-2}$ is on the verge of completion.
This first sample, upon which the results on clustering and large-scale 
structure presented in the next sections are based, has been
constructed to be at least 90\%  complete \cite{Hans98}. Several
external checks, as  
comparisons with independently extracted sets of clusters, support
this figure.  95\% of the 
candidates in this sample are confirmed and observed spectroscopically,
while all redshifts should be measured by the summer of this
year\footnote{Note added in proof: as of July 1999 only 3 clusters
have no measured $z$}.  The distribution on the sky of the REFLEX
clusters in this complete sample is shown in Figure~\ref{aitoff_kp},
while their 3D distribution can be appreciated from the cone diagram of
Figure~\ref{cone_kp}.  From this latter figure we can see how at this
flux limit, the depth of the REFLEX survey is similar to
that of the most recent galaxy surveys, as 2dF the and SDSS.  At the
same time, given the large solid angle of REFLEX, only the SDSS will
be able to explore a comparable volume.   Of course, clusters provide
a coarse--resolution mapping of structures with respect to galaxies,
but it is a price one is happy to pay, as in parallel extremely large
scales can be explored with a reasonable investment of telescope time.

\section{Statistics of Large-Scale Structure}
\label{sec:stat}

In the previous sections we have mostly limited our discussion 
to the observed general characters of the large--scale structure
of the Universe within $z<0.2$, as described by the distribution
of luminous objects. 
The visual appearance of large--scale structures -- while very
interesting {\it per se} -- needs 
to be translated into a quantitative description through the
application of statistical estimators of clustering if one wants
to compare the data to model predictions.

The increased size of galaxy and cluster redshift samples that we have
discussed in the previous sections, has in parallel given the possibility
to produce more and more
accurate estimates of the {\it two--point 
correlation function} and the {\it power spectrum} of the distribution
of these objects.  This has
allowed us to start looking at the {\it details} of the shape of these
functions, in particular on large scales, where we have seen they are
more interesting for the theory.   Such details, if confirmed, can
have profound consequences for our understanding of the origin of
large--scale structure, as I shall try to summarize and discuss in
this section.

\subsection{The Galaxy Two--Point Correlation Function}
\label{sec:xi}

The simplest approach to clustering is to ask how much does it differ
from a uniform distribution at the two--point level, or in other words,
which is the
{\it excess} probability over random to find a galaxy at a separation
$r$ from another galaxy.  This is one way in which the two--point
correlation function \xir can be defined (see \cite{Peebles80} for a more
detailed introduction).   The first estimates of the two--point
correlation function go back to the seventies \cite{Groth-Peebles1977},
but the lack of large redshift samples limited these early analyses
to the {\it angular} correlation function $w(\theta)$.  This is
related to \xir through the {\it Limber equation} \cite{Peebles80}
\begin{equation}
w(\theta) = \int_0^\infty dy\, y^4\, \phi^2 \int_{-\infty}^\infty dx\,
\xi\left(\sqrt{x^2+y^2\theta^2}\right) \,\,\,\,\,   ,
\end{equation}
where $\phi$ is the {\it radial selection function} expected for the
2D survey being analysed. 
 
The basic description of galaxy clustering that emerged from these works
is still valid today on small and intermediate scales:
$w(\theta)$ is well described by a power law $\propto
\theta^{-0.8}$, corresponding to a spatial correlation function
$(r/r_o)^{-\gamma}$, with $r\simeq 5\hmpc$ and $\gamma\simeq -1.8$,
and a break with a rapid
decline to zero around $r\sim 10-20\hmpc$.  As we shall discuss in the
following, modern surveys have significantly improved our knowledge of
the two--point correlation functions especially on scales $>10\hmpc$.
However, before discussing these most recent results, it is important
to briefly describe how galaxy peculiar velocities affect the
observed shape of \xir. 

\subsubsection{REDSHIFT SPACE DISTORTIONS}

The actual detection of true intrinsic deviations of \xis from a power
law is complicated in the analysis of redshift surveys by the effects
induced by 
galaxy peculiar velocities.   Here $s$ is now used to make it explicit
that separations are in reality not measured in true 3D space, but in
{\it redshift space}: what we actually measure when we take the
redshift of a galaxy is the quantity $cz=cz_{\rm true}+v_{\rm pec//}$,
where $v_{\rm pec//}$ is the component of the galaxy peculiar velocity
along the line of sight.  This component, while typically $\sim 100
\kms$ for ``field'' galaxies, can rise above $1000 \kms$ in rich
clusters of galaxies.  This distorts the real--space correlation
function in different ways, depending on the scale.  
The resulting \xis, is in general {\it flatter}
than the real--space \xir.  This is the result of two competing
effects:   
the small--scale pairwise velocity dispersion, mostly dominated by
high--velocity pairs in clusters of galaxies (i.e. those within 
the so--called ``Fingers of God''), damps the amplitude of \xir below
$\sim 3\hmpc$. 
On the other hand, coherent flows towards large--scale structures
enhance the contrast of those structures lying perpendicularly to
the line of sight, thus amplifying \xis in the linear regime (i.e.
above $10-20\hmpc$).  I will not enter here into details on how 
the wealth of information on the dynamics of galaxies contained in
these distortions can be extracted, but limit myself to a discussion
on how to correct them to recover the true shape of \xir.  A more
complete discussion can be found, e.g., in \cite{Fisher94_vel} and
\cite{JAP}.     

Redshift space distortions can be corrected either at a rough
level, through a simple statistical compression of the ``Fingers of God'' 
(e.g. \cite{Ghigna96}), or in a more 
appropriate way by computing the correlation function \xip,
where the separation vector \ss between two objects is split into two
components $r_p$ and $\pi$ related as $s^2 = r_p^2+\pi^2$.  This
two--dimensional correlation function can then be projected along the
line--of--sight direction, to obtain the function 
\begin{equation}
w_p(r_p)=\int_0^\infty \xi(r_p, \pi)\, d\pi \,\,\,\,\, ,
\end{equation}
which is independent of redshift--space distortions.  We have already
encountered this function in Figure~\ref{pp_xi_types} when comparing
the clustering strength of early-- and late--type galaxies in real
space (and thus free of their rather different peculiar velocity
fields).  \wp can either be used to constrain the parameters of a
chosen model for \xir, as e.g. the classical $\xi(r) =
(r/r_o)^{-\gamma}$ \cite{G97}, or inverted through the Abel integral
relations to recover the whole \xir.  

All these problems are obviously absent when one analyses
$w(\theta)$, where on the other hand the strongest uncertainty in the
de--projection lies in the knowledge of the radial selection function.
  
\subsubsection{THE LARGE--SCALE SHAPE OF \xir}

The simple form observed from the first estimates of $w(\theta)$ and
\xir at small separations was consistent with the expectations of
gravitational growth from some initial spectrum of fluctuations (at
the time thought to possibly be a simple {\it white--noise}\
i.e. $\propto k^0$, see next section): as gravity has no
built--in preferential scale, a power law seemed to be a natural
consequence of gravitational clustering (see e.g. \cite{Peebles80}).
However, since then we have understood that plausible initial
conditions are all but a white--noise (see e.g. the first computation
of the linear power spectrum in a Universe dominated by Cold Dark
Matter, \cite{firstCDM}), so that the clustering we measure today with
\x is not just the product of the nonlinear action of gravity.  

\begin{figure}
\centering
\psfig{file=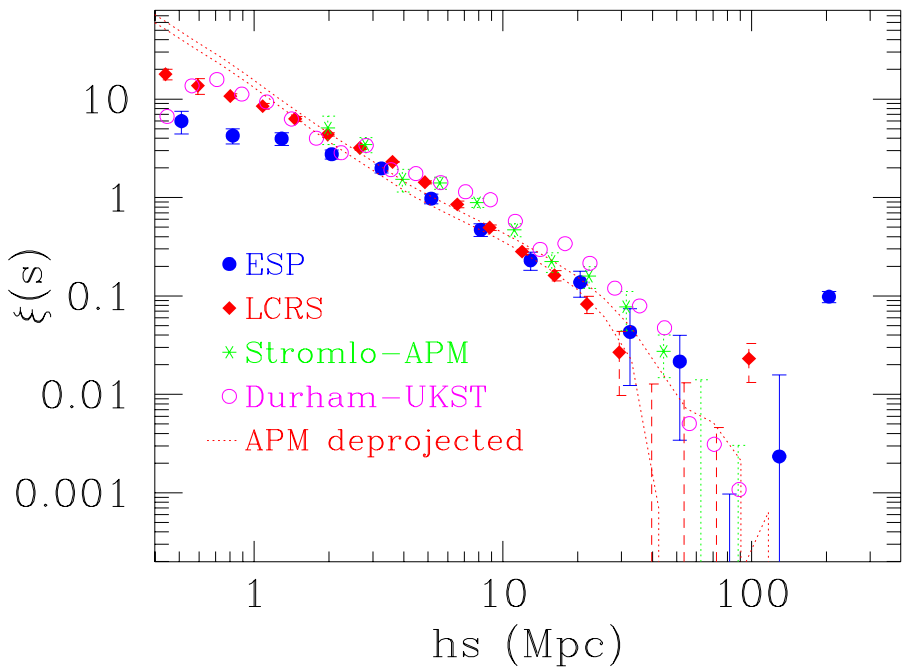,width=13cm}
\caption{Recent estimates of the two--point correlation function of
optically--selected galaxies.  The plot shows results from the ESP 
(Guzzo et al. 1998, 1999), the LCRS (Tucker et al. 1997), the APM-Stromlo 
\cite{Loveday92b} and the Durham-UKST \cite{Ratcliffe et al. 1996}
surveys.}
\label{xi-surveys}
\end{figure}

In Figure~\ref{xi-surveys}, I have plotted the estimates of \xis for
the ESP \cite{Guzzo98, Guzzo99}, the LCRS \cite{Tucker}, the
Stromlo--APM (Loveday et al. 1992b),  
and the Durham--UKST \cite{Ratcliffe et al. 1996} surveys.  These samples
represent a selection of data that should offer the best compromise
between depth and angular aperture, thus maximising the ability to
sample large scales.  In addition, the dotted lines show a plot of
\xir obtained through de--projection of the angular $w(\theta)$ from
the APM galaxy catalogue \cite{Baugh96} under two different
assumptions about galaxy clustering evolution and thus selection function.

We clearly see how, within some scatter among
different surveys, \xis remains positive to separations of $50\hmpc$
or larger.  Keeping in mind how the variance in galaxy counts is $\sim
1$ around $8\hmpc$, we can conclude that a significant range of scales
over which we measure positive clustering is still in the linear or
quasi-linear regime.   

The global form below $5-10\hmpc$ is still well
described by a power law: the slope is very close to the classical
$-1.8$ for the APM \xir, which is in real space, while it is flatter
for all the redshift--space measures due to the suppression by
peculiar velocities discussed above.
Above $\sim 5\hmpc$ there is unanimous evidence for more power than
expected by a simple extrapolation of the small--scale slope.   This
``bump'' or ``shoulder'' is evident both in the APM \xir and in the
redshift--space measures, implying that it is not an effect of the
expected redshift--space amplification by coherent flows
\cite{Kaiser87}.   We found clear early evidence for this excess when
studying clustering in the Perseus--Pisces survey \cite{G91},
and realised that it was present already in the published CfA1 data
(as also noticed in \cite{Dekel_Aarseth84}).  At the time we suggested
that it was and indication for a steep 
power spectrum $P(k) \propto k^{-2.2}$ on large scales.   Further
theoretical modeling \cite{Branchini94} and the new direct
measures of \pk\ in real space from the APM survey \cite{APM_pk}, 
confirmed that indeed there is a significant change in \xir
around $r\simeq 3-5\hmpc$.  It was natural to interpret this as a
consequence of the transition between the strongly nonlinear
clustering regime at small separations, to a quasi--linear regime on
larger scales.   In \S\ref{sec:pk} we shall come back to this point
while discussing directly the observed shape of \pk.

\subsection{The Clustering of Clusters}

We have seen that clusters of galaxies represent a powerful tracer of
structure on the largest possible scales.  Their clustering can be also
quantified at the simplest level through the two--point correlation
function.  The classic estimate of \xis for Abell clusters
\cite{Bahcall_Soneira83} showed that the cluster--cluster 
correlation function is also well described by a power law, with a slope
apparently similar to that of galaxies, but a correlation length 
about 4 times larger.  In reality, due to the limited size of the
sample, the original fit was performed {\it imposing} a slope $\gamma=1.8$,
and therefore it was not really a measure of the functional shape of
cluster--cluster correlations.
Nevertheless, the fit was good enough, and it became generally accepted that 
the cluster-cluster correlation function has a the same slope as galaxies,
$\gamma=1.8$, but larger amplitude (see e.g. \cite{Kaiser84}), that is
$\xi_{cc}(r)\simeq A\cdot \xi_{gg}(r)$. In fact, this statement {\it
could not} rigorously be true if 
a simple statistical amplification mechanism, as then suggested by
Kaiser \cite{Kaiser84}, were the origin of the different amplitude: clusters
trace scales $>10\hmpc$, i.e. cover mostly fluctuations that are in the
quasi--linear or linear regime, and it would have been a rather
strange conspiracy, that their slope were the same that galaxies
display on scales between 0.1 and $5 \hmpc$, where clustering is
highly nonlinear. 
\begin{figure}
\centering
\psfig{file=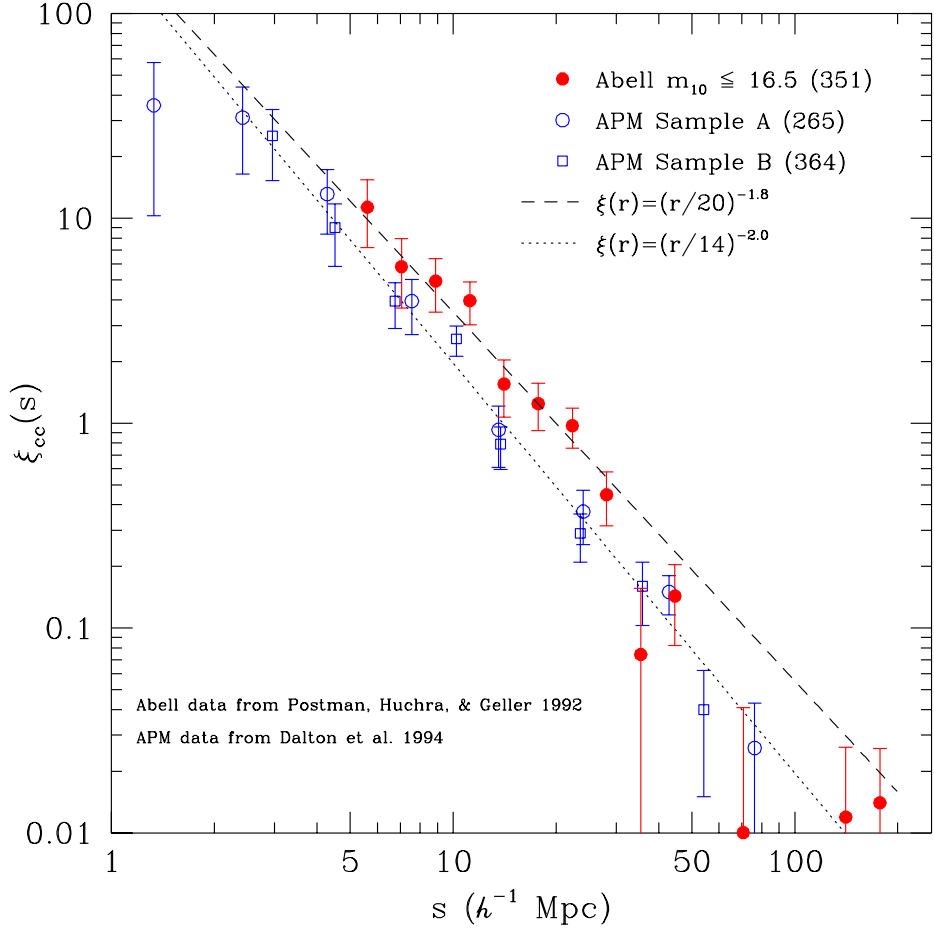,width=13cm}
\caption{The cluster--cluster correlation function optically--selected
clusters from the Abell and APM catalogues, reproduced from the review
by Postman \cite{Postman98}. }
\label{xi-abell}
\end{figure}

The basic problem was that the galaxy correlation function was not known
accurately enough on large scales, as to provide a meaningful
comparison.  The situation has fortunately improved significantly since
then.  We have just reviewed the significant progress made in our knowledge of
the galaxy correlation function.  In parallel, new cluster samples
have been constructed, such as the EDCC \cite{Collins95} and APM \cite{Dalton}
automatically selected cluster catalogues, and the quality and number
of redshifts available for Abell clusters have substantially
increased.

Although I will not enter into details concerning optically--selected
clusters, in Figure~\ref{xi-abell} I have reproduced a plot from
\cite{Postman98}, showing an up--to--date comparison of the  
cluster--cluster correlation functions of both an Abell sample
\cite{PHG92}, and a sample of APM clusters \cite{Dalton}.  While
comparison is presented with two possible power laws, the data clearly
show a break from these simple models around $50\hmpc$, much in the
same way as galaxies do on a similar scale.  See \cite{Postman98} for
more details. 

\begin{figure}
\centering
\psfig{file=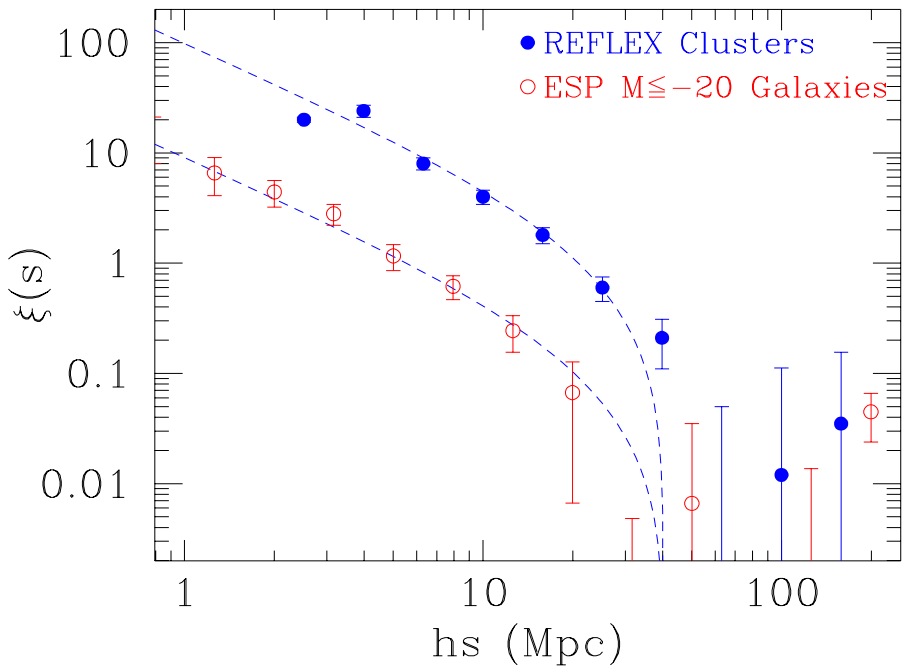,width=13cm}
\caption{Comparison of the two--point correlation functions of
REFLEX clusters \cite{xi_REFLEX1} and ESP galaxies \cite{Guzzo99}. 
The top dashed line is the Fourier transform of a simple
phenomenological fit of the REFLEX power spectrum with a
double--power--law model.  The bottom one is the same after scaling by
an arbitrary bias factor of $b_c^2=(3.3)^2$.  The agreement in shape
between galaxies and clusters is remarkable.  At the same time, both
galaxies and clusters show an indication, after the breakdown around
$r\simeq 50 \hmpc$, for more positive power on scales exceeding $100
\hmpc$.} \label{xi-ESP-REFLEX}
\end{figure}

In section \S\ref{sec:reflex} I argued that X-ray selection is
the best way  to select homogeneous samples of clusters with
well--defined physical criteria out to large redshifts.  In particular, X-ray
luminosity is a parameter that is much more closely related to mass than the
somewhat loosely defined {\it richness}, used to characterise 
optically--selected clusters.  For this reason, model predictions for
the clustering of massive objects, can be more easily and safely translated in
terms of observable quantities as luminosities and fluxes, than in the
case of galaxies \cite{XBACS}.  At a simpler observational level, it
is particularly interesting to compare $\xi_{cc}$ for X-ray selected clusters
to that of galaxies, as we do in  Figure~\ref{xi-ESP-REFLEX}
\cite{REFLEX_mess2}. 
This figure shows a preliminary estimate of \xicc from the
flux--limited REFLEX survey \cite{xi_REFLEX1}, compared to the galaxy--galaxy
correlation function from two volume--limited subsamples of the ESP
survey \cite{Guzzo99} \footnote{The use of
volume--limited samples is to be preferred when discussing the shape of
\xis.  Estimates of \xis from whole magnitude--limited surveys are
normally subject to weighting schemes, as e.g. the so--called J3
minimum--variance weighting, which, while allowing a better sampling
of very large scales, can affect the globale shape of \xis (Guzzo et
al. 1999) .  Volume--limited samples are much better defined in terms
of the properties of the galaxies they include, containing only
objects with luminosity above a well--defined threshold.}.   
The dashed line on top of the cluster points is the Fourier
transform of the power spectrum of REFLEX clusters (computed
independently, see next section) while the bottom line has been
scaled down by an  
arbitrary factor $b_{cg}^2=(3.3)^2$, so as to overlap the galaxy
points. The agreement 
between the shapes of the cluster and galaxy correlation functions is
remarkable.  Here we also see how a proper functional description of
the shape is not just a simple power law.  The one shown here is the
Fourier transform of the simple phenomenological shape for \pk
suggested by Peacock \cite{JAP}.

The result shown in Figure~\ref{xi-ESP-REFLEX} is a powerful confirmation of a
simple linear bias model between galaxies and clusters of galaxies,
analogous to eq.~(\ref{sigma}), as first suggested by Kaiser
\cite{Kaiser84} (see also \cite{Mo_White96} for a more recent
refinement).  Indeed, considering the relationship between \xir and
the variance within a top--hat sphere of radius $r$
\begin{equation}
\sigma^2(r) = {1\over V_r}\int_{V_r} \xi(x) d^3x \,\,\,\,  ,
\end{equation}
eq.~(\ref{sigma}) implies that the cluster and galaxy correlation
functions obey to the relation
\begin{equation}
\xi_{cc}(r) = b_{cg}^2 \xi_{gg}(r)\,\,\,\,  ,
\end{equation}
where the relative bias factor $b_{cg}$ is related to the typical mass
of the clusters considered \cite{Kaiser84}.  This kind of
investigation can be generalised to the study of the dependence of the
correlation length on the sample limiting X--ray luminosity, for which
also model predictions can be quite specific \cite{Mo_White96}.  This
will also be an important output of the REFLEX survey \cite{xi_REFLEX1}. 

\subsection{The Power Spectrum}
\label{sec:pk}

The Fourier transform of the correlation function is the power spectrum
P(k)
\begin{equation}
P(k) = 4\pi \int_0^\infty \xi(r) {\sin(kr) \over kr} r^2 dr \,\,\,\,\, ,
\end{equation}
which describes the distribution of power among different wavevectors
or {\it modes} $k=2\pi/\lambda$
once we decompose the fluctuation field $\delta = \delta\rho/\rho$ over
the Fourier basis \cite{JAP}. 

The amount of information contained in $P(k)$ is thus formally the
same yielded by the correlation function.  The estimates of $P(k)$ or \xir
from redshift surveys, however, are affected in different ways by
uncertainties introduced, for example, by the poor knowledge of the
mean density (in which case the power spectrum is to be preferred),
or by the shape of the survey volume (whose effect is usually more
easily treated when computing \xir rather than \pk).  Useful
references for learning more about this topic are 
\cite{Ed_Valencia}, \cite{Strauss97} and \cite{JAP}, where further
directions can be found to specific technical papers.
One practical benefit of the description of clustering in Fourier
space through \pk\ is that for fluctuations of very
long spatial wavelength ($\lambda > 100 \hmpc$), where \x is  dangerously
close to zero and errors easily make the measured values fluctuate
around it, $P(k)$ is on the contrary very large.
Around these scales, most models predict the power spectrum to have a
maximum, which reflects the size of the horizon at the epoch of
matter--radiation equivalence.   

Indeed, comparison of observations to the theory is in
principle easier and more direct using \pk.  First, models are usually
specified in terms of a linear \pk, which is the result of the action
of the specific {\it transfer function} of the model on a {\it
primordial spectrum}, usually assumed to be of the so--called
Harrison--Zel'dovic scale--invariant form $\propto k^1$, which is also
the kind of spectrum most naturally produced in inflationary scenarios
(see e.g. \cite{Col-Luc} for more details).   In addition, k--modes in
Fourier space are statistically independent (a part from the
convolution effects due to the {\it window function} of the survey,
see below), and direct $\chi^2$ comparisons to models is feasible,
which is in principle not the case when the correlation function is
analysed (see e.g. \cite{Fisher94_xi}).

However, not everything is better with power spectra.  Redshift surveys
are all but cubes (i.e. what would be optimal for a Fourier
plane--wave decomposition), and their geometrical shape affects the
measured power, so that what we really measure is the quantity
\begin{equation}
\tilde P(\kvec) = \int d\kvec^\prime \left|W(\kvec-\kvec^\prime)\right|^2\,
P(\kvec^\prime)\, +\, SN \,\,\,\,\,\,	.
\end{equation}
The measured power spectrum is therefore a {\it convolution} of the
true $P(\kvec)$ with the square modulus of the {\it 
window function} $\left|W(\kvec)\right|^2 $, that is the Fourier
transform of the survey volume, plus an additional shot--noise term.
While the shot--noise contribution $SN$ is easily corrected for, the
recovery of the true \pk\ 
necessarily involves a delicate de--convolution operation in $k$
space.  While for nearly tridimensional surveys (as IRAS-based
surveys \cite{Fisher93, Feldman94, Tadros_Efstathiou95}, or the
CfA2--SSRS2 \cite{CfA2_SSRS2_PK}, Stromlo--APM \cite{APM-Stromlo-PK},
Durham--UKST \cite{Hoyle99}, and REFLEX surveys), the effect of the
window function is mostly negligible, for nearly two--dimensional surveys as
the LCRS or, even worse, the ESP, its effect is dramatic to very small
wavelengths.   The key point is that for slice surveys like ESP, the
window function is very anisotropic, in particular it is extremely large
along the direction perpendicular to the main plane of the survey.
When the final 
estimate of \pk\ is computed by averaging over the whole $4\pi$ solid
angle, this anisotropy brings contributions from different $\kvec$'s
into the same averaged $k$ bin.   It is important to keep these
limitations in mind when one compares estimates of \pk\ from different
surveys as we shall do here.  A comprehensive discussion
on different estimators for \pk\ and how to take these effects into
account can be found in \cite{Tegmark_pk}.

A different approach for dealing with surveys with peculiar shapes is
otherwise that suggested by Vogeley \& Szalay \cite{KL}, using the
so--called Karhunen--Lo\`eve transform.  Rather than trying to correct
the effect of the window function over the plane waves of the Fourier
basis, the idea is to find a different set of orthonormal eigenvectors
which are optimal given the survey geometry.  The interesting
quantities, as e.g. \pk, are then projected on this basis, both for
the data and for the models, and comparison is performed through a
maximum likelihood analysis. Application of this method has been so
far limited only to the 2D case \cite{KL}.   A first application to
the REFLEX data \cite{Schuecker} is yielding promising results. 

\subsubsection{THE POWER SPECTRUM OF THE GALAXY DISTRIBUTION}

\begin{figure}
\centering
\psfig{file=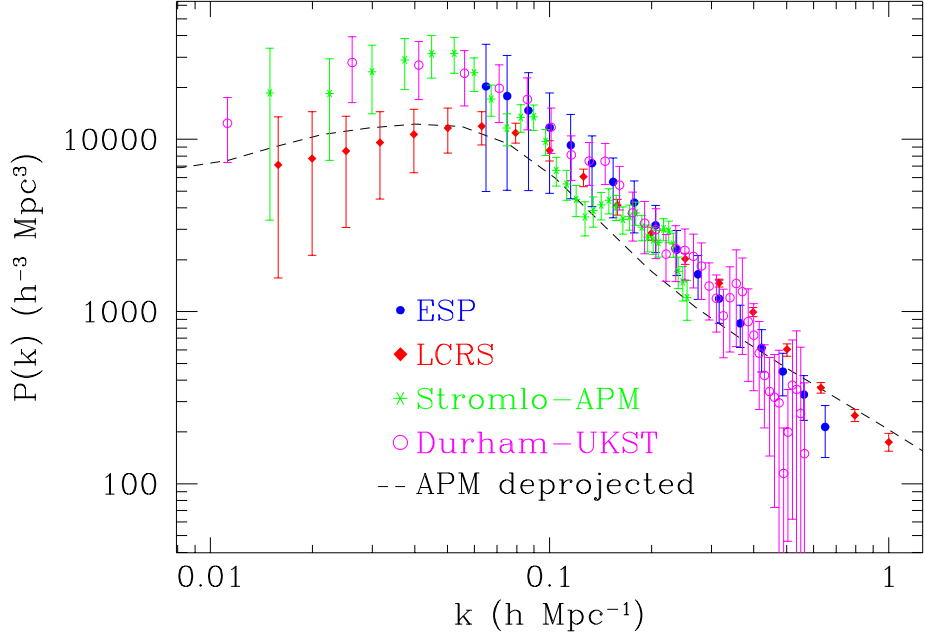,width=13cm}
\caption{A non--exhaustive compilation of most recent estimates of the
power spectrum of galaxy clustering, from four of the largest
available redshift surveys of optically--selected galaxies (ESP
\cite{Carretti99}; LCRS \cite{pk_LCRS}; Stromlo-APM
\cite{APM-Stromlo-PK}; Durham-UKST \cite{Hoyle99}), compared to that
deprojected (and therefore in real space), from the 2D APM galaxy
survey \cite{APM_pk}.}
\label{pk_gal}
\end{figure}

In Figure~\ref{pk_gal}, I have plotted the estimates of \pk\ for the
same surveys as given in Figure~\ref{xi-surveys}\footnote{One further 
notable estimate, not shown here, has been recently produced from the
IRAS--based PSCz survey, and can be found in \cite{PSCZ_pk}}.  The
four data sets allow me 
also to make a comparison of estimates from relatively tridimensional surveys
(Stromlo--APM, Durham-UKST), to more bidimensional samples as the LCRS
and ESP, the latter being in practice a single thin slice cut through
the galaxy distribution.  This means that the effect of the window
function (and the need of a proper correction) on these data sets is
very different.  In addition, three of the four samples are selected 
in exactly the same photometric band, the blue--green $b_J$ (two, ESP
and Durham--UKST are even constructed from the same catalogue, the
EDSGC), the only exception being the $r$-band selected LCRS. This has the
positive effect of reducing the relative biasing between the different
samples, although some effect is possibly still present due to the
different luminosity ranges covered.

In the same figure, I have also plotted (dashed line) \pk\ as
reconstructed from the projected angular clustering of the APM galaxy
catalogue \cite{APM_pk}.  This is clearly the only
estimate which is free of redshift--space distortions.  The effect of
these is shown
in particular by the slope above $\sim 0.3\kmpc$: an increased
slope in real space (dashed line) corresponds to a stronger damping by
peculiar velocities, diluting the apparent clustering observed in redshift
space (all points).

The first impression from this comparison is that to first order there
is quite a good agreement across the different samples.  The general trend is
that of a well--defined power law range between $\sim 0.08$ and $\sim
0.3 \kmpc$, with a slope around $k^{-2}$.  Note that despite these
samples are rather similar in terms of their galaxy properties, a
minimal level of relative biasing might be present because of the
relative weight of faint and bright objects. The only two samples that
should in principle display the same amplitude are the ESP and the
Durham--UKST surveys, which are both selected from the EDSGC catalogue.
In fact, their \pk\ are 
practically identical over a good range of $k$'s.  On one hand, the
Durham--UKST \pk\ becomes rather noisy at small scales, due to the
sparse sampling strategy of this survey (less sparse than the
Stromlo--APM, though).  On the other hand, the ESP \pk\ performs well
on small scales, but on large scales it has to be limited to $k>0.06\kmpc$, 
below which the effect of its nasty window function cannot be
deconvolved appropriately.  In fact, the very good agreement with the
Durham--UKST down to fairly small $k$'s is a very encouraging
indication of the quality of the deconvolution procedure performed by
the authors\cite{Carretti99}.

We also note how the LCRS power spectrum tends to be flatter and of
lower amplitude around the tentative turnover displayed by the
Durham--UKST and Stromlo--APM spectra.  Also at large $k$'s, where ESP
and Durham--UKST are significantly damped by small--scale pairwise
velocities, the LCRS \pk\ seems to be less affected by this distortion.
The same trend is also visible in the correlation function plot of
Figure~\ref{xi-surveys}. 

Recalling the discussion of \S~\ref{sec:xi} on the shoulder 
observed in the two--point correlation function above $\sim 5 \hmpc$,
here we can see clearly the same effect in \pk\ by looking at the
real--space power spectrum from the APM 
catalogue.  The slope of the APM \pk\ above $0.3\kmpc$ is $\sim
k^{-1.2}$, corresponding to the small--scale clustering regime (in
real space!) where $\gamma\simeq -1.8$.  Below this scale, \pk
steepens to $\sim k^{-2}$, and this is what produces the excess power
in \xis on large scales.   

Peacock \cite{jap97} applied the sophisticated linear
reconstruction machinery by Hamilton and collaborators
\cite{Hamilton} to the whole \pk\ 
from the APM catalogue (and to another real--space estimate from the 
the IRAS-QDOT survey \cite{Saunders_QDOT}) and concluded that the
shape was indeed consistent with a linear power spectrum
characterised by a steep slope ($\sim k^{-2.2}$).  This is the same
value of linear slope originally suggested in \cite{G91} to
explain the observed large--scales shape of \xis in the 
CfA1 and Perseus--Pisces redshift surveys, and
confirms the early speculation that the observed change in slope of
\xir is a manifestation of the transition from the quasi--linear to the
strongly nonlinear clustering regime. 

\subsubsection{THE POWER SPECTRUM FROM CLUSTERS}

Also for measuring the power spectrum, clusters of galaxies
offer the most efficient alternative to galaxies, given their ability
to sample very large volumes.
A first preliminary estimate of $P(k)$ from the REFLEX survey is shown
in Figure~\ref{pk_clusters} \cite{Schuecker}, compared to the power
spectra of the largest redshift sample available for Abell clusters
\cite{Retzlaff98}, and that of the APM automatic optically--selected
clusters \cite{pk_APM_clusters}.  This is a conservative measure, based
on only 188 of the nearly 460 clusters with redshifts that will form
the final REFLEX sample, using a Fourier box of $400\hmpc$
comoving side.  This was done to avoid possible spurious fluctuations
due to the incomplete sampling between the Northern and the Southern
galactic sides of the survey.  At the time of writing this review,
virtually all clusters in the survey have been observed
spectroscopically. A new measure of $P(k)$ in a box of $\sim
1000\hmpc$ side (i.e. using all clusters within $z\sim0.2$, where
the survey is complete), should be produced by the
end of 1999\footnote{Note added in proof: a new estimate of \pk\
within such a volume obtained just before completing the final version
of this paper, can be found in \cite{Dunk}}.
Despite the point at smaller $k$'s is not significant, 
the turnover around $k\simeq 0.05 \kmpc$ is shown to be significant at
the $3\sigma$ level by a Karhunen-Lo\'eve \cite{KL}
Maximum Likelihood 
analysis\footnote{That is, projecting the data and a phenomenological
form for \pk, with two power laws connected at a scale $x_c$, over the
best basis of eigenvectors found for the REFLEX
geometry\cite{Schuecker}.}.   The analysis of the whole survey
should also allow to put more serious constraints on the detailed
shape of $P(k)$ around the turnover.  

\subsection{Features in the Power Spectrum}

\begin{figure}
\centering
\psfig{file=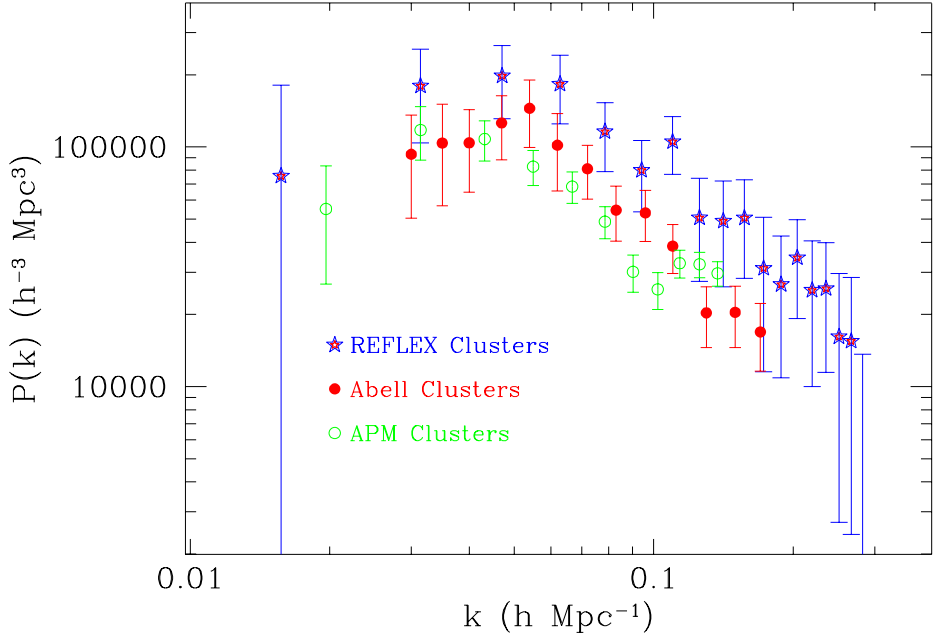,width=13cm}
\caption{The power spectrum of the clustering of clusters, as measured
from three different cluster surveys: a preliminary subsample of 188
clusters from the X-ray selected REFLEX survey \cite{Schuecker}, and
redshift samples from the Abell \cite{Retzlaff98}, and APM
\cite{pk_APM_clusters} cluster 
catalogues.  Note that a systematic 
difference in amplitude among these surveys is expected as they sample
different mass thresholds, and are therefore characterised by
different bias values.}  
\label{pk_clusters}
\end{figure}

During the last few years, evidence has indeed been accumulating
that the peak of the power spectrum could be rather sharp, perhaps
characterised by an extra feature (with respect to smooth traditional models)
around its maximum.   Einasto and collaborators \cite{Einasto_peak}
analysed the distribution of Abell clusters, finding evidence
for a sharp peak around $k\simeq 0.05\kmpc$. Although their result was
questioned by other workers because of the possible incompleteness in
the sample and the way $P(k)$ was estimated, the presence of such a
feature was confirmed by a more conservative reanalysis of Abell data
\cite{Retzlaff98}, where a slightly less pronounced but still
significant peak is found, as we show in Figure~\ref{pk_clusters}.   
The peak is indeed evident in the Abell data, while the preliminary
REFLEX sample is not yet able to put serious constraints on the shape
around the turnover.   Note also the very good agreement of the slopes 
of the three power spectra above $k\simeq 0.5\kmpc$, and
at the same time the shift in amplitude of the three samples.  The
latter is a clear manifestation, again, of the different bias of the
three samples.  An X-ray selected sample as REFLEX allows for a more
direct link to the typical mass of the objects we are looking at (see
\cite{Borgani_Guzzo2000}, for a more extended discussion of these points).

If we also recall that  evidence for excess power around $\sim 100
\hmpc$ was provided by a 2D power spectrum analysis of the LCRS slices
\cite{Landy_2D}, we cannot avoid being amused by the consistency
in the peak scale to which these separate measures point, i.e., 
consistently between 100 and $150\hmpc$.  This is remarkably close to the 
``periodicity'' scale revealed by Broadhurst and collaborators
(\cite{BEKS}, BEKS hereafter), in the analysis
of their 1--dimensional pencil--beam surveys towards the galactic poles.
This latter result has certainly been one of the most exciting findings of 
this decade in the study of large--scale structure.  The authors merged
together two deep redshift surveys of redshifts performed
independently in the direction the two galactic poles over a small
field of $\sim 0.7$ degrees, exploring a total 1--dimensional 
baseline of $\sim 2000\hmpc$.  The resulting galaxy distribution showed
a surprising regularity of ``spikes''.  The visual impression was
quantified  and confirmed by a 1D correlation and power spectrum
analysis, that clearly indicated a preferential ``fluctuation'' at $128\hmpc$. 

This result originated significant controversy. It was suggested that it
could just be an aliasing of power due to the small size of the beam, that
projected power from small to large scales \cite{Kaiser_Peacock}.  
On the other hand, the reality of the effect was supported by independent 
observations showing how the more nearby peaks detected in the pencil beam
were coincident with known, real large--scale structures, as the Great
Wall or the Sculptor supercluster (e.g. \cite{EDCC92}).  Further
pencil beams in different directions (e.g. \cite{Ettori97}), and a
denser sampling around the original pointings \cite{beks99} also show
that, yes,  this direction is somewhat special, but 
only in the sense that here the effect is maximised.  This is
what one would statistically expect if there is indeed a distribution
of typical ``cell'' sizes around a characteristic dimension
\cite{voronoi}.  The peak observed in the 3D 
power spectrum of Abell clusters around the same scale is further
suggesting that the origin of the BEKS periodicity lies indeed in a
specific feature in the 3D power distribution around this wavelength.  

Note also in Figures~\ref{xi-surveys}, \ref{xi-abell}
and~\ref{xi-ESP-REFLEX} the 
behaviour of the two--point correlation functions on very large
scales, for both galaxies and clusters. Although 
the binning of \xis is very coarse at these separations,
there is a hint that \xis becomes positive again around
$150-200\hmpc$.  This seems to be common to nearly all surveys,
independently of their geometry (slice or 3D surveys), the kind of
tracer (galaxy or clusters), and the estimator used.  As can be
readily seen by Fourier 
transforming a ``standard'' $P(k)$ (e.g., a CDM shape \cite{BBKS}),
this damped oscillation of \xis cannot be reproduced if 
\pk\ has a smooth turnover around its maximum, and seems to be a
further hint for a sharp peak.  A similar oscillation in the
correlation function was claimed for Abell clusters \cite{Oscill_xi}, and
interpreted as evidence for a sharp feature in $P(k)$.  

\begin{figure}
\centering
\psfig{file=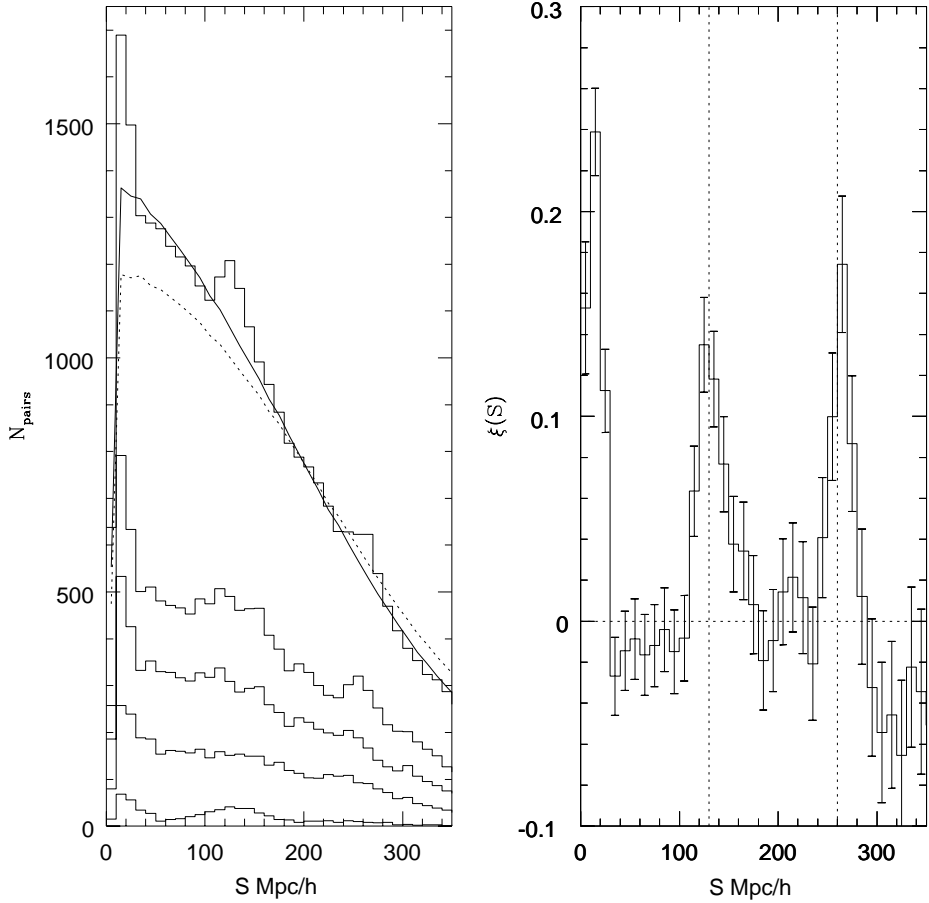,width=13cm}
\caption{The one--dimensional clustering of Lyman--break galaxies,
from \cite{Br_Jaffe}. Left: the co--added pair counts from
5 fields at $z\sim 3$ (top histogram), compared to the expectations
from a randomly distributed sample with the same selection function
(solid line).  Right: the 1D correlation function along the line of
sight, showing the clear pattern with $\sim 130\hmpc$ periodicity
(using $\Omega=0.2$).  See \cite{Br_Jaffe} for details.}
\label{Ly-peaks}
\end{figure}

At the time of writing this review (Spring '99), one very interesting
piece of evidence has been provided along the same lines by Broadhurst
and Jaffe \cite{Br_Jaffe}, who analyse the redshift distribution of the
high--redshift samples of Lyman--break selected galaxies by Steidel
and collaborators \cite{Steidel98}.  As shown in
Figure~\ref{Ly-peaks}, they find again the same
effect detected at smaller redshift, i.e. the emergence of a preferred
clustering scale.  One important consequence is that the co--moving
scale of the peak in the power spectrum measured locally ($\sim
130\hmpc$), can be used as a standard stick to provide a constrain on
the combination of $\Omega_{M}$ and $\Omega_\Lambda$:
$48\Omega_M-15\Omega_\lambda\simeq 10.5$, which for a flat Universe 
($\Omega_{M} + \Omega_\Lambda=1$), gives $\Omega_{M}=0.4\pm0.1$.

The convergence of so many independent observations seems to have left
little doubt, in my opinion, that the observed characteristic scale is
real, and is telling us something important about the properties of
our Universe.   Indeed, on the theory side 
there has been considerable interest in the
recent literature about the possibile relation of this peak to baryonic
acoustic features produced within the last scattering surface
at $z\sim 1000$, when the Cosmic Microwave Background (CMB) radiation
originated.  Eisenstein and collaborators \cite{Eisenstein}, show
however how a large baryon 
fraction ($\Omega_b/\Omega_o \simeq 0.3$ or larger), and a rather {\it
ad hoc} combination of parameters (as e.g. a ``blue" tilt of the
primordial spectrum) are required to match the observed Abell peak, while
at the same time being consistent with the cluster
abundance\footnote{Note added in proof: a similar model is found to
provide a good description of the most recent estimate of \pk\ from the
REFLEX data \cite{Dunk}}.   
More dramatically, it is worrying to see that no realistic CDM
parameter combination is capable to account for the excess power (the
``shoulder") we were discussing in section~\ref{sec:xi}, that is
displayed by virtually all modern surveys \cite{Meiksin}.

Even without considering the existence of extreme features, therefore
there seems to be a general difficulty for the ``standard'' theory to
explain the 
detailed shape of $P(k)$, now that the data are becoming of higher and
higher quality in the linear regime.  Especially when 
one tries matching the power spectrum implied by the growing amount
of CMB  anisotropy experiments to that
displayed by the clustering of luminous matter, problems seem to be
unavoidable.   According, for example, to Silk \& Gawiser
\cite{Silk_Gawiser} ``If 
the data are accepted as being mostly free of systematics and {\it
ad hoc} additions to the primordial power spectrum are avoided, there
is no acceptable model for large-scale structure."  Once again, a
major uncertainty prevents any firm statement from being made: are we
allowed to compare the power spectrum derived from the distribution of light to
that derved from the mass (CMB), through a simple linear bias scaling?
Clearly, a 
scale--dependent bias would add room for any detailed match between
models and the data, but without a solid physical basis would also add
an unpleasant {\it ad hoc} taste to the whole picture.  In 
addition, the extremely linear relation between the correlation
functions of galaxies and clusters that we have shown in
Figure~\ref{xi-ESP-REFLEX}, seems to suggest that at least above $\sim
5 \hmpc$ the galaxy and mass distributions are linked by a simple
linear bias.

\section{Evolution of Large--Scale Structure}

\begin{figure}
\centering
\psfig{file=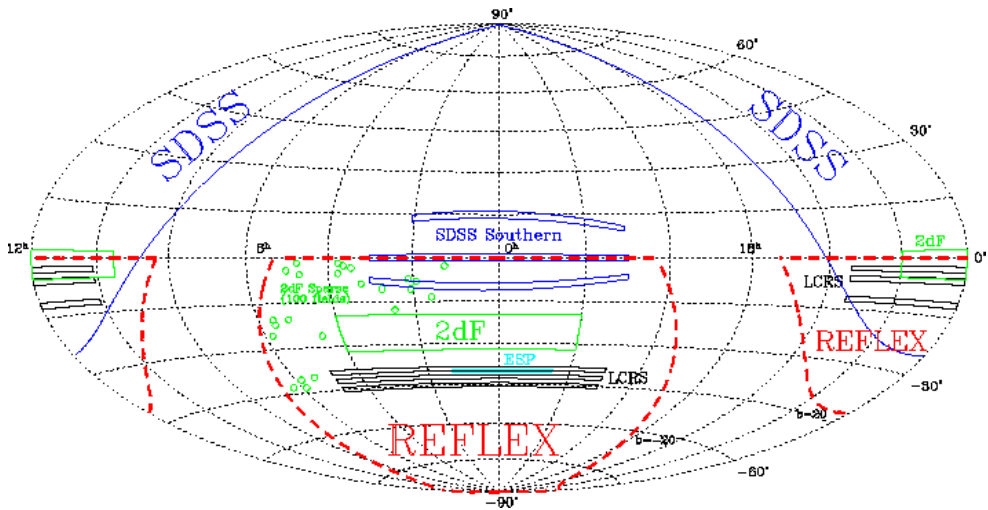,width=13cm}
\caption{Pictorial view in Aitoff projection of the areas covered by
the major galaxy and cluster redshift surveys discussed in the text.}
\label{aitoff_surveys}
\end{figure}
Figure~\ref{aitoff_surveys} gives a pictorial view of the sky
coverage by wide--angle redshift surveys of galaxies and clusters
that have mapped or are in the course of mapping the ``local''
Universe to $z\sim0.2$\footnote{This is an up-to-date version of a plot
which already appeared in \cite{G96}, and \cite{Strauss97}.}.   
All these surveys have been discussed in some detail in the previous
sections.

While considerable effort will still be necessary to complete the
coverage both in photometry and redshift of this volume of
the Universe\footnote{For example, it seems natural to think
that at some point a multicolour survey as the SDSS will necessarily
have to be extended to the southern sky.}, massive redshift surveys are
already being planned to study large--scale structure at redshifts of
the order of unity. 

Deep redshift surveys, as demonstrated by the results from the
clustering of Lyman--break galaxies discussed in the previous section,
give in principle the possibility to trace the
growth of structure, which represents a formidable test for theories:
a successful model must be able not only to reproduce the final
picture, but also all the snapshots of the ``cosmic movie'' at
different epochs.

However, when we start exploring a baseline in time comparable to
galaxy evolutionary times, we not only have the problem of
understanding how our galaxy tracers map the underlying mass density
field, but also to know the way they evolve, i.e. we need a full
comprehension of the bias function $b(R,t)$.   
A successful theory must therefore
provide two different ingredients: 1) a cosmological background, i.e.  a
linear power spectrum, depending on H$_o$, $\Omega$ and to the kind
of dark--matter particles dominating the mass density, including a
possible curvature contribution by a cosmological constant 
$\Lambda$ ; 2) a recipe to convert the mass and the growth history of
``objects'' (i.e. dark 
matter halos), forming during the gravitational evolution from the
chosen initial conditions, into radiation with some spectral
distribution, to be ``observed'' and compared to the real data.  This
is the arena of semi--analytical galaxy formation models (see the
review by Kauffmann, these proceedings).  Here halos obtained through
analytical ``merging trees'' (e.g. \cite{Baugh_models, Somerville}), or
from purely gravitational N--body simulations \cite{Governato98} are
``lite--up'' through as realistic 
as possible analytical recipes.   This seems at present the best we
can do to properly attach some ``flesh'' on the evolving ``bones'' of
large--scale structure.   It is self--evident that the
inherent complication of realistically treating the dissipative part
of this process (e.g. cooling, star formation, stellar evolution,
supernova feedback -- just to mention a few of the branches into which
the problem of galaxy formation needs to be decomposed to become
tractable), represents the weakest aspect of this machinery, and makes
the use of galaxies as tracers of the evolution of large--scale
structure a rather tricky game.

\subsection{Non--Evolving Tracers?}

Could we perhaps find a way to select some kind of tracer that does not
evolve (or at least such that its evolution is weak and simple to
understand), out to some cosmologically significant redshift?
Clearly, if 
we push this redshift limit to some indefinitely high value, this
cannot be true for any class of objects, as sooner or later we shall
hit the epoch of their major formation.   However, we could hope to
isolate specific redshift ranges over which the growth of structure is
significant, while the intrinsic mean spectrophotometric properties of
such tracers remain the same.   

To select such objects, detailed {\it colour} information,
i.e. multi--band photometry, is fundamental.  This implies that future
deep galaxy redshift surveys will necessary have to be based on
photometric catalogues covering possibly from the $U$ band in the
ultraviolet to the $K$ band in the infrared.  This is the case for the
VIRMOS deep survey, that I shall briefly describe in \S\ref{sec:virmos}.   

The power of colour selection to isolate classes of objects within
specific redshift ranges has been extraordinarily demonstrated by the
Lyman--break selection technique \cite{Steidel98}.  These
strategies, which base their power on the existence of ``breaks'' in
the spectrum of galaxies, tend clearly to select galaxies for which
these features are particularly prominent.  The technique of Steidel
and collaborators selects preferentially galaxies at $z\sim 3$ or
$z\sim 4$ (depending whether $U$--band or $B$--band drop-outs are
selected), with strong star formation, that enhances the Lyman break
at 912\ \AA.  While Lyman--break galaxies represent a class of
objects whose properties would have changed  significantly by the present
time, it is possible to use similar techniques to select ``steady''
objects as early--type
galaxies, for which another  spectral break, that at 4000 \AA\, is
particularly prominent,  within a well--defined redshift interval.  

For example, Iovino and collaborators \cite{Iovino_xi} have been able, using 
Schmidt plates in three
bands,  $b_J$, $R$, and $I$, to construct a sample of several
thousands early--type galaxies
within the redshift range $0.3 < z < 0.55$, and measure their
clustering.  They find a correlation length $r_o\sim 6 \hmpc$, to be
confronted to the value of more than $ 8\hmpc$ at the present epoch (see
Figure~\ref{pp_xi_types}).   This is one of the cleanest measures of
the evolution of large--scale structure presently available: through
the selection of a slowly evolving class of objects we try to
circumvent our ignorance about how galaxies in general form and
evolve. In this way, the measured difference in the clustering
strength should only reflect the growth of fluctuations, and therefore
the cosmological model. 
We have seen already how the application of this technique to the
unique multi--band photometric catalogue of SDSS will create in a
similar way a volume--limited sample of 100,000 early--type galaxies
with measured redshift.   

\subsection{Large Redshift Surveys to $z\sim 1$ and Beyond}
\label{sec:virmos}

The exploration to $z<1$ was pioneered at the beginning of the nineties
by deep pencil beam surveys as BEKS, and the Canada-France Redshift
Survey \cite{CFRS}.  A summary of clustering results from deep surveys,
together with the most recent advances related to the Lyman--break
selected galaxies can be found in \cite{Steidel98}.  Here I would like
to spend a few words describing a large survey in an advanced
stage of preparation, that promises to enlarge by two orders of
magnitudes the number of available redshifts at $z\sim1$: the VIRMOS
Deep Survey.

This survey has its origin in the construction of two spectrographs for 
visible and IR light (VIMOS and NIRMOS), for the UT3 and UT4
telescopes of the  ESO VLT, which is presently carried out by a
consortium of French and Italian institutes led by O. Le F\`evre
\cite{Olivier_IAP}.  120 nights of guaranteed time will be spent with
these two instruments, starting in late 2000, to perform a 
redshift survey of 150,000 galaxies in the redshift range $0.3<z<5$.

The broad goal of the VIRMOS deep survey is a comprehensive study of
the formation and evolution of galaxies, with particular emphasis on
the evolution of 
the luminosity function, star-formation rate, clustering and the
fundamental plane.    To reach these goals, the following observations
are planned:
\begin{itemize}
\item Spectroscopy of more than 100,000 galaxies with magnitude
$I_{AB}<22$ observed at low resolution over four fields with a total
area of $15-20$ deg$^2$. 
\item Spectroscopy of $\sim 50,000$ galaxies with magnitude $22<I_{AB}<24$
observed at low resolution over an area of 1-2 deg$^2$.
\item  Spectroscopy of $\sim 10,000$ galaxies at higher resolution for study
of the fundamental plane of galaxies
\item  Ultra-deep spectroscopy using an Integral Field Unit (IFU),
(that is, a pack of 6400 fibres over a $1\times1$ arcmin$^2$ field), of about
1000 galaxies to $I_{AB}\simeq 26$. 
\end{itemize}

\begin{figure}
\centering
\psfig{file=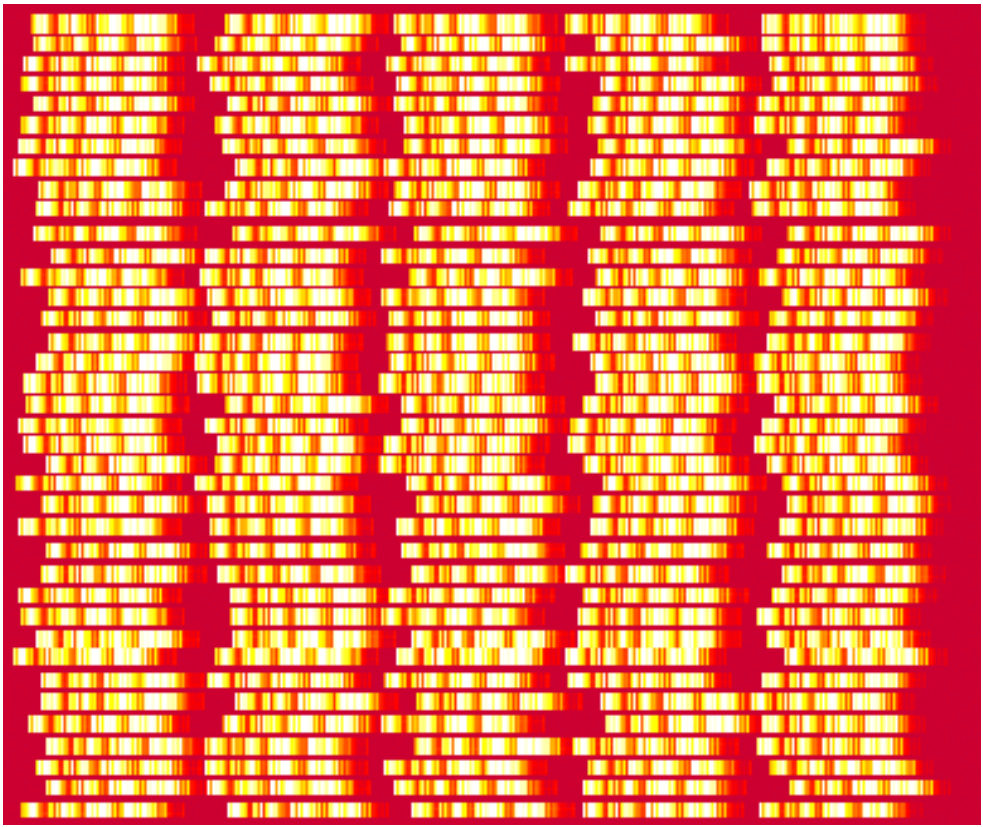,width=13cm}
\caption{A simulated CCD frame as it will be produced by one of the four
quadrants of the VIMOS spectrograph, showing 1/4 of the nearly 800
spectra that can be observed at once in the maximum multiplexing
configuration.} 
\label{virmos-slits}
\end{figure}

The total numbers listed above will be made possible by the enormous
multiplexing gain of the spectrographs: VIMOS in particular, will have the
ability to collect simultaneously, in low resolution mode, nearly 800
spectra over a total field of $14\times 14$ arcmin$^2$.   This field of view
is split into four quadrants, to reduce the size of the optical elements 
required.   Figure~\ref{virmos-slits} reproduces the simulated
appearance of the image from one of the four CCD detectors, with
nearly 200 spectra packed over the available area. 
The spectroscopic targets for the survey with
VIMOS an NIRMOS are being selected from a UBVRIK imaging campaign that
is currently in progress using the CFHT, ESO-NTT, ESO-2.2m and CTIO-4m
telescopes.  
%

\subsection{X-ray Clusters as Tracers of High--z Structure}

Given our still poor knowledge of galaxy formation and evolution, 
and on the contrary the relative simplicity of the physycs involved in
the X-ray emission from clusters of galaxies, X-ray selected cluster 
have the potential to become one of the best tools for tracing large--scale
structure at high redshifts \cite{Borgani_Guzzo2000}.
The last few years have seen a number
of studies on the evolution of the cluster abundance and of
their X-ray luminosity function (e.g. \cite{Rosati98}).
These works have shown that the abundance of clusters with 
$L < 
10^{44}$ h$^{-2}$ erg s$^{-1} \sim L^*$ 
does not seem to evolve significantly
between the present epoch and $z\sim 1$.  Although some uncertainties
on how to relate this to the underlying evolution of the mass function
still remain, these observations, together with the easiness in
detecting X-ray clusters to high redshifts, qualifies them as
excellent tracers for studying the evolution of large-scale structure.

The kind of flux limits that it is necessary to reach, to be able to
detect at $z\sim 1$ objects with such luminosities, are
comparable to the depth of ROSAT pointed observations, i.e. $\sim
10^{-14} \, erg\, s^{-1} \, cm^{-2}$.   However, to be able to map
large-scale structure at high redshifts, we shall need a survey
covering an area  significantly larger than the $\sim 100$ deg$^2$ typical
of the serendipitous surveys constructed from the ROSAT PSPC archive,
and in particular much less sparse than these. 
Such an X-ray survey has been recently proposed 
as the science drive for the construction of a wide-field X-ray
satellite (\cite{wfxt}).  Crucial for such a survey
instrument is a novel design of the X-ray optics \cite{Burrows} which
are able to obtain stable point--spread function of 
$\sim 10$ arcsec over a 1-degree diameter field, in contrast to the
rapidly degrading resolution of classical Wolter--type telescopes
\cite{Citterio}.  A prototype of such innovative mirrors has
been already constructed and successfully tested in the Merate labs of
OAB.  A more detailed discussion of the use of X-ray clusters as
tracers of the evolution of structure, as a scientific case for future
wide--angle deep X-ray surveys will be found in \cite{Borgani_Guzzo2000}.

\section{Conclusions}

Looking at the various topics addressed in this paper, we can probably
imagine the future development of this field during the first 10 years
of the new Millennium as possibly moving along two directions.

One one side, the completion of the 2dF and in particular of the Sloan
Digital Sky Survey will produce profound progress in the knowledge of
many important statistical aspects of large--scale structure at $z\sim
0$.  First
of all, it is clear that something very interesting is hidden in the
exact shape of the power spectrum near the expected turnover.  While we seem
finally to detect such turnover with some significance from the best
samples now 
available, the data are still too poor on scales of $500-1000\hmpc$ to
sample it with sufficient resolution in $k$
space.   This feature is more clearly detected when high--threshold
objects as clusters of galaxies or Lyman--break galaxies are analysed.  
Larger samples of clusters of galaxies will be therefore essential:
the REFLEX survey will be pushed down in flux by about a factor of two
within the next two years, which will bring the number of clusters
in the sample up to $\sim 800$, probing a volume extending to $z\sim
0.3$.  At the same time, the sample of red luminous galaxies from the
SDSS seems to represent the most promising data set for exploring the
scales around the peak of $P(k)$ in sufficient detail.  Also, 
the question on whether these features in the distribution of objects
are partly a product of biasing, and therefore cannot be transferred
to the matter 
power spectrum, will find an answer when detailed CMB anisotropy
experiments will explore such scales.   There seem to be all premises
that the Planck Surveyor  (see e.g. Efstathiou, these proceedings)
will be able to provide this information to a high level of accuracy.

At the same time, 10--meter class telescopes equipped with a new
generation of spectrographs give the opportunity to extend systematic
studies of large samples of galaxies to redshifts above unity.  I have
briefly described the largest of these projects, the VLT--VIRMOS deep
survey.  Similar surveys with smaller sizes are being planned by
different groups around the World, as newer instrumentation on very
large telescopes commences operation.  One example that I did not
have time to discuss here is the Deimos spectrograph, under construction
for the Keck II telescope \cite{deimos}.

The study of very large volumes of the Universe at high redshifts,
however, will have to resort to using tracers which can be both more
efficient and easier to understand than galaxies.   X-ray selected
clusters of galaxies provide this opportunity.  Consequently, it would
be highly significant if within the next
$\sim10$ years a large, possibly all--sky, survey in the X-ray band to a flux
about 2 orders of magnitude fainter than the ROSAT All-Sky Survey
($\sim 10^{-14}$ erg s$^{-1}$ cm$^{-2}$) be performed.  Concrete proposals
in this direction have recently been presented \cite{wfxt}, and will
hopefully become reality in the near future.

\section*{Acknowledgments}

I would like to thank the organizers of the 19$^{th}$ Texas Symposium
on Relativistic Astrophysics for inviting me to give this review talk.
I thank all my collaborators in the large--scale structure
projects I have described throughout this paper, particularly
Hans B\"ohringer, Guido Chincarini, Chris Collins, Olivier Le F\'evre,
Peter Schuecker, Paolo Vettolani and Elena Zucca.  A special
acknowledgment goes to Stefano Borgani for a number of enlightening
discussions on the use of X-ray clusters as tracers of large--scale
structure, for teaching me a lot about theoretical cosmology and for
reading the manuscript. Thanks are due also to Fiona Hoyle,  
Helen Tadros, Matthew Colless and Huan Lin for providing electronic
versions of their results and to Silvia Quarello for useful discussions.  I
am also grateful to Rychard Bouwens for carefully reading an early
version of the manuscript and providing useful comments. 

Finally, special thanks should go to my wife and my kids, for their
patience during the hours spent writing this review during weekends,
and for alleviating the difficulty of finishing it, through
their company and background playing.

\section*{References}


\end{document}